\documentclass[%
 reprint,
superscriptaddress,
 amsmath,amssymb,
 aps,
pra,
]{revtex4-2}

\usepackage{lineno}
\usepackage{braket}
\usepackage{graphicx}
\usepackage{dcolumn}
\usepackage{bm}
\usepackage{hyperref}

\hypersetup{colorlinks=true, citecolor=blue, urlcolor=blue, linkcolor=blue}

\begin{document}

\preprint{APS/123-QED}

\title{High-Rate Point-to-Multipoint Quantum Key Distribution using Coherent States}

\author{Yiming Bian}
 \affiliation{%
 State Key Laboratory of Information Photonics and Optical Communications, School of Electronic Engineering, Beijing University of Posts and Telecommunications, Beijing, 100876, China
}%
\author{Yichen Zhang}%
 \email{zhangyc@bupt.edu.cn}
 \affiliation{%
 State Key Laboratory of Information Photonics and Optical Communications, School of Electronic Engineering, Beijing University of Posts and Telecommunications, Beijing, 100876, China
}%
\author{Chao Zhou}%
 \affiliation{%
 State Key Laboratory of Advanced Optical Communication Systems and Networks, School of Electronics, and Center for Quantum Information Technology, Peking University, Beijing 100871, China
}%
\author{Song Yu}%
 \affiliation{%
 State Key Laboratory of Information Photonics and Optical Communications, School of Electronic Engineering, Beijing University of Posts and Telecommunications, Beijing, 100876, China
}%
\author{Zhengyu Li}%
 \email{lizhengyu2@huawei.com}
 \affiliation{%
 Huawei Technologies Co., Ltd., Shenzhen 518129, China
}%
\author{Hong Guo}%
 \email{hongguo@pku.edu.cn}
 \affiliation{%
 State Key Laboratory of Advanced Optical Communication Systems and Networks, School of Electronics, and Center for Quantum Information Technology, Peking University, Beijing 100871, China
}%

\date{\today}

\begin{abstract}
    Quantum key distribution (QKD) which enables information-theoretically security is now heading towards quantum secure networks. It requires high-performance and cost-effective protocols while increasing the number of users. Unfortunately, qubit-implemented protocols only allow one receiver to respond to the prepared signal at a time, thus cannot support multiple users natively and well satisfy the network demands. Here, we show a `protocol solution' using continuous-variable quantum information. A coherent-state point-to-multipoint protocol is proposed to simultaneously support multiple independent QKD links between a single transmitter and massive receivers. Every prepared coherent state is measured by all receivers to generate raw keys, then processed with a secure and high-efficient key distillation method to remove the correlations between different QKD links. It can achieve remarkably high key rates even with a hundred of access points and shows the potential improvement of two orders of magnitude. This scheme is a promising step towards a high-rate multi-user solution in a scalable quantum secure network.

\end{abstract}

\maketitle

\begin{figure}
    \centering
    \includegraphics[width=8 cm]{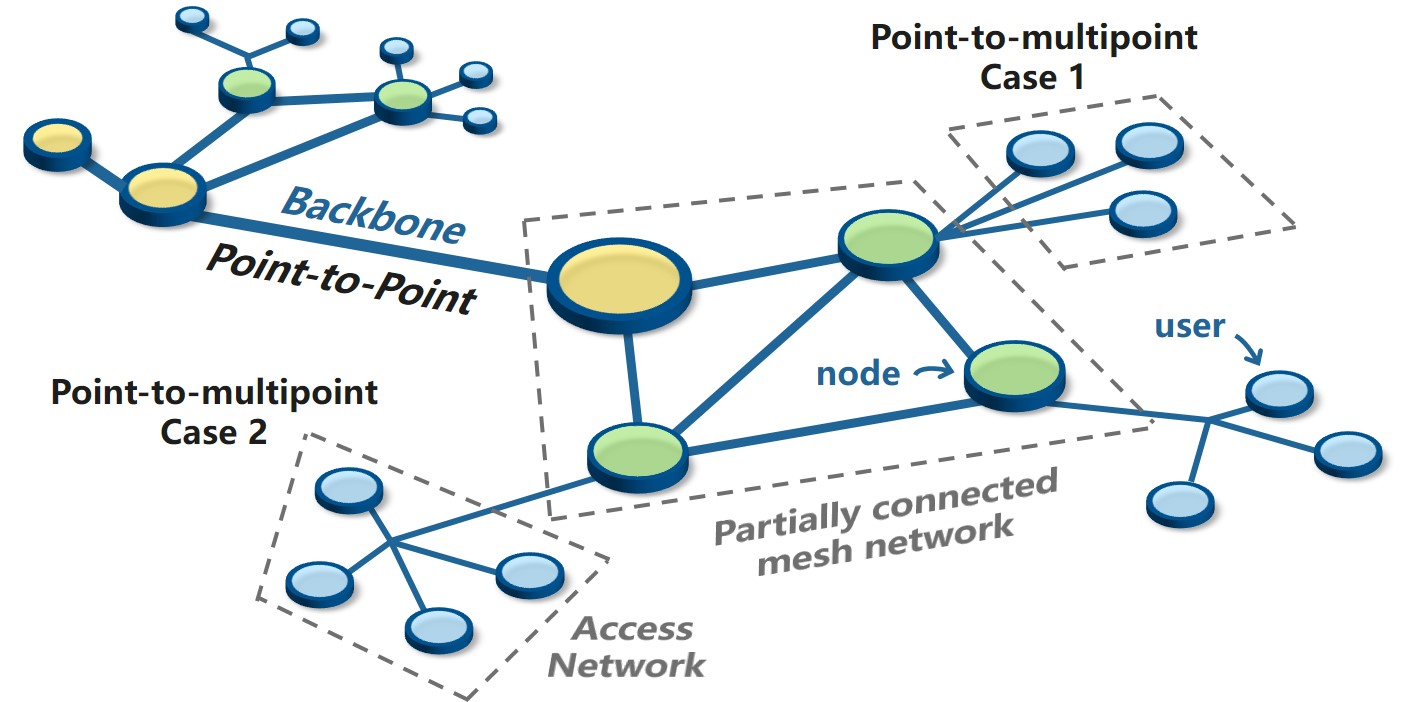}
    \caption{ \textbf{An efficient quantum secure network topology.}  The yellow nodes represent the main network node of a metropolitan network, which are connected together by a backbone network. The green nodes are the secondary network nodes of a metropolitan network, connected with a partially connected mesh network. The blue nodes are the end users connected with the network nodes.}\label{QSNet}
\end{figure}

Quantum key distribution (QKD) \cite{bennet1984quantum, AdvInQC, PTPQKDRMV2020, portmann2022security} allows secret key generation between two distant parties using the correlation established by the quantum process. It can be realized by encoding the secret information on quantum states within a finite or infinite Hilbert space, corresponding to the discrete variable and continuous variable (CV) protocols \cite{GG02Nature,GaussianQuantumInformation,lam2013continuous} respectively. The discrete variable QKD has experienced a long period of development and can support a rather long distance \cite{PTPQKDTFNat2018, 800kmTFQKD}, while CV-QKD is advantageous in the compatibility with classical optical communications and high key rate within metropolitan distances \cite{CvExp80km2012,CVMDIYork,CVQKD50km,CvExpSOICVQKD2019,CvExp202kmPRL,CVNC2022,CvExp5GRep}.
In pace with the maturity of the point-to-point links, QKD is developing towards networking, namely quantum secure network.
It has been realized from the metropolitan-area network \cite{QCnetVienna2009, QCNetToyoko, CambridgeQN, HuweiQN} to the large scale wide-area network \cite{Micius,QCnetNature2021}, even with space-to-ground links \cite{PTPQKDSatellite2017,QCnetNature2021,Micius}. Most of the networks can be implemented with a topology shown in Fig. \ref{QSNet}, where point-to-point links connect the main nodes, and point-to-multipoint (PTMP) links connect the users with the nodes together.

However, QKD can hardly support multiple users in protocol layer. Until now, QKD protocols are mostly designed for two parties, which can well support a point-to-point QKD link but cannot natively support the interconnection of multiple end users. Especially for the discrete variable protocols, the information encoded on a single photon can only be used by a pair of two users to build correlation, which limits the possibility of supporting multiple users and increases the detection costs. The existing metropolitan \cite{46node} and access networks \cite{QNetNature2013,QAN2015} based on the two-user QKD protocols inevitably face the limitation of functionality, connectivity and scalability \cite{joshi2020trusted}.
The lack of multi-user QKD protocols has seriously hindered the development of quantum secure network.

\begin{figure*}
    \centering
    \includegraphics[width=18 cm]{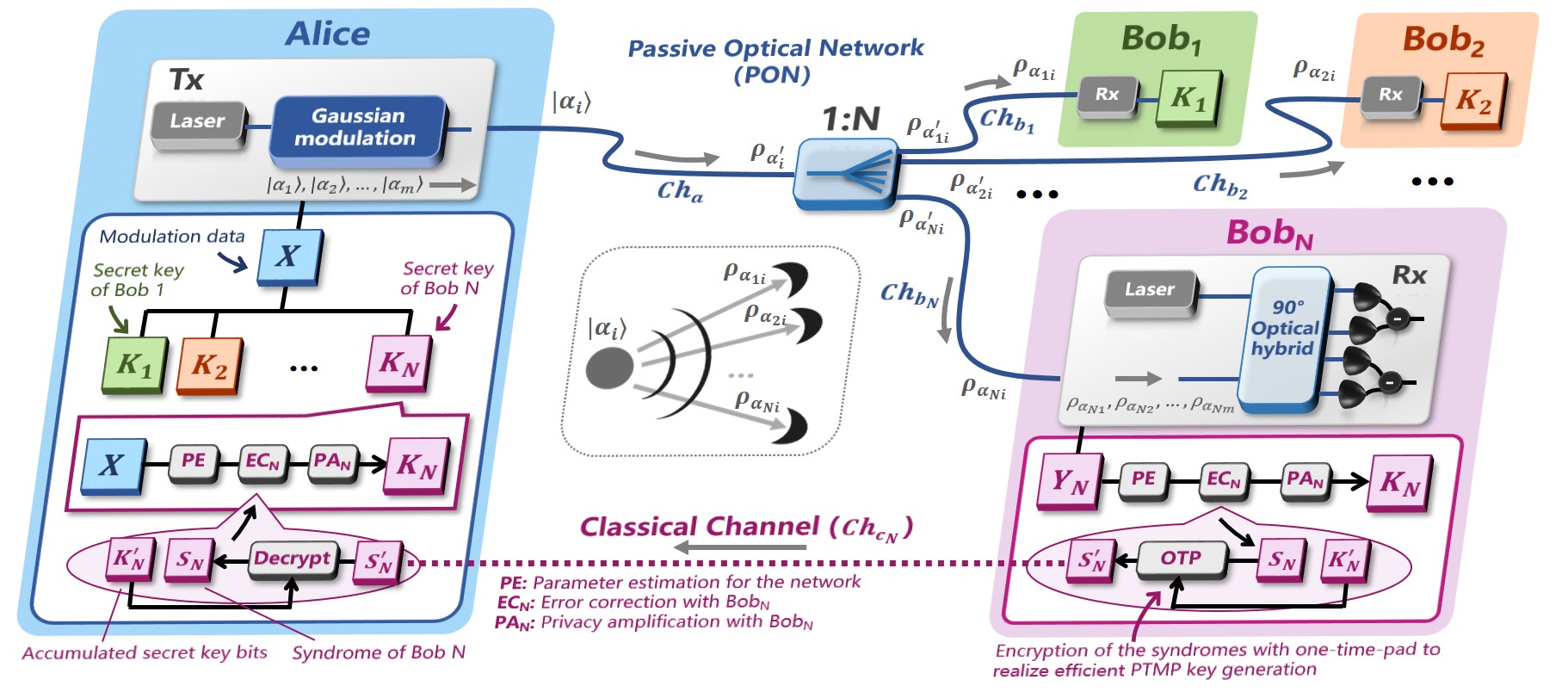}
    \caption{\textbf{Structure of the protocol.} Alice prepares quantum states $\ket{\alpha_1}, \ket{\alpha_2}, ..., \ket{\alpha_m}$ and sends them to each Bob, the modulation data is saved as $X$, and the detection data of different Bobs are $Y_1,Y_2,...,Y_N$.
    Here, when $\ket{\alpha_i}$ is prepared, Bob$_1$, Bob$_2$, $. . .$ , Bob$_N$ get $\rho_{{\alpha_{1i}}}, \rho_{{\alpha_{2i}}}, ... ,\rho_{{\alpha_{Ni}}}$ respectively.
    Then, Alice performs post-processing with each Bob to generate $N$ independent secret key bits strings with $N$  Bobs based on $X$.
    Especially, the syndromes produced by Bobs during error correction is designed to be encrypted with one-time-pad (OTP) for the efficient key generation of all Bobs. For Bob$_N$, the syndrome $S_N$ is encrypted with the accumulated secret key bits $K_{N}'$ between him and Alice. Note that, here we present the structure of a local local oscillator system \cite{CvExpLLO2015, CvExpLLO2015-2}, but the protocol also works with a traditional system where the local oscillator is prepared by the transmitter \cite{CvExp80km2012,CvExp202kmPRL}.
    }\label{Structure}
\end{figure*}

In this article, we propose a PTMP QKD protocol that aims to support multiple users in a most common network structure, PTMP connection. Correlations between multiple parities are simultaneously built based on CV system \cite{GG02PRL,NSPRL,CVSecure2006,CVSecure2006second,CVSecure2014, DMCVJapan}, with one party (Alice) preparing the coherent states and multiple parties (Bobs) measuring them. The transmission of the states can be realized by a structure of the optical power splitter. Similar to a process of the broadcast, each state evolves into multiple states through different paths, which are finally measured independently by different Bobs. This process builds the correlation between Alice and each Bob concurrently, also makes each Bobs correlated. In security analysis, these correlations are used to get a tighter analysis of the protocol, helping to achieve high key rate. During data processing, the correlations between different Bobs are protected from using by eavesdropper, and removed by privacy amplification to make the secret key independent and secure. In this way, QKD is firstly enabled to support multiple users in protocol layer, where each prepared quantum state can be measured and processed by multiple parties for generating independent secret key bits, significantly enhancing the protocol efficiency.

Our protocol can support up to 128 receivers with a transmission distance over 125 km, the secret key rate of each Bob is higher than $10^{-6}$ bit/pulse at 100 km, and $10^{-3}$ bit/pulse at 10 km.
In a practical access network case, the secret key rate for each user can still reach more than 54 kbps within 25 km with 128 users, higher than that of the access network with point-to-point protocols accessing 8 users in a shorter distance \cite{QNetNature2013,QAN2015}.
For a network with 32 users, the transmission distance can exceed 40 km, making it suitable for the access and metropolitan area network.
The downstream networking strategy here benefits the practical deployment, since the massive calculation of error-correction decoding is finished in the transmitter side, making end user side as simple as possible.  Besides, the network structure and the coherent detection also provides the potential of being integrated in classical networks, providing a promising way to realize a high-rate and cost-effective quantum secure network.

~\\
\textbf{Results}
~\\
\textbf{PTMP protocol.}
~\\

\begin{table*}
    \caption{PTMP QKD protocol.}
    \label{Table1}
    \begin{tabular}{m{17 cm}}
        \hline
        1. \textbf{State preparation}: Alice prepares quantum states $\ket{\alpha_1}, \ket{\alpha_2}, ..., \ket{\alpha_M}$ with Gaussian modulation. Her modulation data is saved as $X$, including the modulation data for $x$ and $p$ quadrature. Normally, $X=[(x_1^A,p_1^A),(x_2^A,p_2^A),...,(x_M^A,p_M^A)]$. \\

        2. \textbf{Measurement}: Alice sends the states to the quantum channel, normally a PON. All Bobs (Bob 1,2, $...$, $N$) receive the quantum states transformed from the states prepared by Alice. For one state $\ket{\alpha_i}$ from Alice, the corresponding states received by $N$ Bobs are $\rho_{\alpha_{1i}}$,\ $\rho_{\alpha_{2i}}$,\ …,\ $\rho_{\alpha_{Ni}}$. They respectively use heterodyne detection to measure the $x$ and $p$ quadrature of their received states. The detection data from all Bobs are saved as $Y_1, \ Y_2, \ ...,\ Y_N$. Note that for a no-switching scheme, both $x$ and $p$ quadratures are saved, thus basis sifting is not required. Normally, the detection data of Bob $i$ is $Y_i=[(x_{i1}^B,p_{i1}^B),(x_{i2}^B,p_{i2}^B),...,(x_{iM}^B,p_{iM}^B)]$. \\

        3. \textbf{Parameter estimation}: Alice discloses part of the modulation data, and all Bobs disclose their corresponding detection data. By calculating the variance and covariance, a covariance matrix, $\gamma_{AB_1B_2\ldots B_N}$, can be estimated, which reflects the characteristics of the whole protocol. The remaining data after the disclosing is $X',\ Y_1', \ Y_2', \ ... , \ Y_N' $. Based on the covariance matrix $\gamma_{AB_1B_2\ldots B_N}$, the upper bound of the information known by Eve, the correlation between different Bobs, and the secret key rate of each Bob are estimated.\\

        4.  \textbf{Error correction}: Alice communicates with each Bob respectively to transform $X'$ and $\ Y_1', \ Y_2', \ ... , \ Y_N' $ into $N$ same bit strings $D_1,D_2,...,D_N$. Normally, they convert their continuous-variable data to discrete-variable form, $(D_1^A,D_1^B),\ (D_2^A,D_2^B),\ ...,\ (D_N^A,D_N^B)$, and then calculate the syndromes respectively.
        In reverse reconciliation, for Bob $i$, he calculates the syndrome $S_i$ based on his data $D_i^B$. Then he uses the accumulated secret key bits $K_i'$ shared between him and Alice to encrypt his syndrome with one-time-pad, and transmits the encrypted syndrome to Alice through a classical channel.
        Alice uses the corresponding secret key bits $K_i'$ to decrypt the syndrome, and corrects her data with it. If successful, Alice and Bob $i$ will share the same bit string $D_i$. \\

        5. \textbf{Privacy amplification}:
        By utilizing the universal hashing functions, Alice and each Bob remove the information may known by Eve and the correlation between the other Bobs to get $N$ independent secret key bit strings $K_1,\ K_2,\ ...,\ K_N$. \\
        \hline

    \end{tabular}
\end{table*}

The prepare-and-measure scheme of the PTMP protocol is shown in Fig. \ref{Structure}, and detailed in Box \ref{Table1}.
Each coherent state prepared by Alice can make all Bobs respond, after coherent detection, the correlations between Alice and all Bobs are established.
These correlations contributes to the parameter estimation, since  the more correlation between the legitimate parties, the less information Eve can get from the channel.
We can construct a covariance matrix containing all trusted modes ($\hat{A}$ and $\hat{B}_1$, $\hat{B}_2$,..., $\hat{B}_N$) to evaluate the situation of the channels tightly.
In this way, the sensitivity of the protocol to the number of the receivers is suppressed, since when more receivers access, though the loss introduced by the optical power splitter increases accordingly, the modes contains by the covariance matrix increases as well, providing more information for security analysis.

After that, the correlations between Bobs should be removed to make the secret key between Alice and each Bob independent.
Here, different Bobs' detection data are correlated because the signals detected by different Bobs evolve from the same quantum state, though the noise introduced during the state transmission and the shot noise make different Bobs' detection data different.
We remark that, when designing the protocol, we hope the correlations between different Bobs can be easily removed and affect the protocol as little as possible. Thus, each  Bobs are asked to perform independent measurement and data processing without cooperation with the others, which avoids the accumulation of the other Bobs' knowledge on one Bob, contributing to the removal of Bobs' correlation. This is studied quantitatively in security analysis.

Before removing the correlation between Bobs with privacy amplification, their correlations also affects the error correction.
During error correction, each Bob sends their syndromes to Alice ($S_1,\  S_2  ,...,\ S_N$), and Alice corrects her data $N$ times with different syndromes to share same but different key strings with each different Bob respectively.
However, the correlations between Bobs may make $S_i$ leak the information about  Bob$_j$ ($i \neq j$).
If the syndromes are disclosed as in one-way protocols \cite{PostProcess2004,PostProcess2008}, Eve may process them  collectively to obtain more information about each Bob. Thus, in PTMP protocol, the syndromes are secretly transmitted with one time pad (OTP) to reduce the impact of Bobs' correlations on  error correction.
In next part, we show that, in protocol layer, the encrypted transmission of syndromes will not cause additional loss of secret key, making  the error correction efficient, and Bobs' correlation mentioned before can be removed with negligible costs.

~\\
\textbf{Security analysis.}
~\\
\begin{figure}
    \includegraphics[width=9 cm]{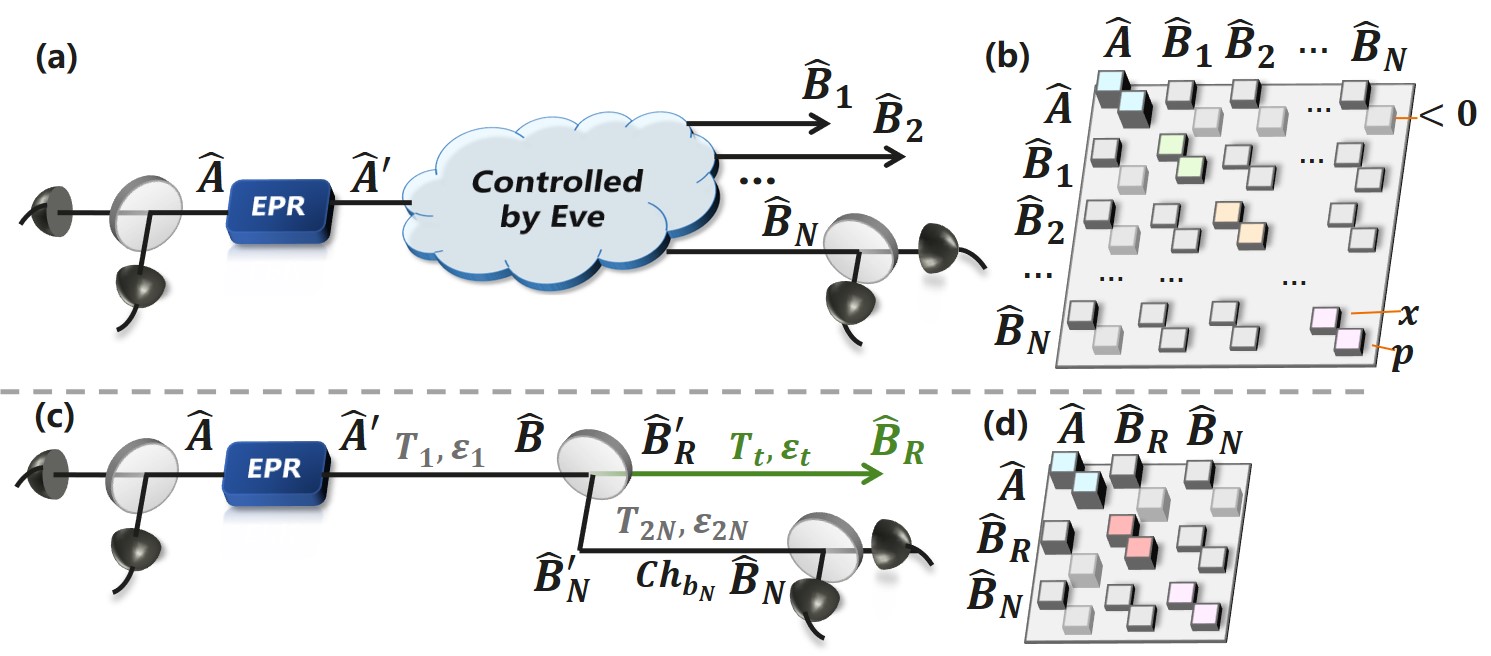}
    \caption{\label{EBSmall}
    \textbf{Entanglement-based (EB) scheme.}
    (a)  The general EB scheme, where Alice performs heterodyne detection to detect the mode $A$, the other mode $A'$  is sent to Bobs through a structure controlled by Eve. Bob$_N$ performs heterodyne detection. The security analysis doesn't depends on the structure of the channel. (b) A simplified scheme when the structure of the network is the same as that in Fig. \ref{Structure}. (c) The covariance matrix $\gamma_{AB_1B_2,...,B_N}$ for the general EB scheme. (d) The covariance matrix $\gamma_{{A}{B}_R{B}_N}$ for the simplified scheme. The covariance matrix represents the  variance of the mode and the covariance between each modes, in $x$ and $p$ quadratures. $T$ and $\varepsilon$ represents the transmittance and excess noise in the quantum channel.
    }
\end{figure}
The security analysis is established on the entanglement-based scheme shown in Fig. \ref{EBSmall} (a), to maintain generality, the structure of the channels is not specified.
In reverse  reconciliation, we can get the secret key rate between Alice and Bob$_N$ as
\begin{equation}
    \label{eq1}
    \begin{split}
        K_N= \beta I\left({A}:{B}_N\right)-\max{\left\{\max_i{I({B}_N:{B}_i)},\chi_{{B}_N{E}}\right\}}.
    \end{split}
\end{equation}
Here, $\beta$ is  the reconciliation efficiency, $I\left({X}:{Y}\right)$ is the classical mutual information, and $\chi_{{B}_N{E}}$ represents the Holevo bound \cite{holevo1973bounds}.

Eve and the other Bobs can be seen as $N$ adversaries relative to Alice and Bob$_N$. Since Eve is an illegal party, we don't restrict her behaviors. Assuming she can fully control the channels of the network and purify the whole system, with the extremality  of  Gaussian  states \cite{wolf2006extremality}, her knowledge is bounded by Holevo bound to present the worst-case
\begin{equation}
    \chi_{B_NE}=S(\rho_{{A}{B}_1{B}_2...{B}_N})-S(\rho^{m_{{B}_N}}_{{A}{B}_1{B}_2...{B}_{N-1}}).
\end{equation}
Here, $S(\rho)$ is the von-Neumann entropy, which can be calculated with the covariance matrix $\gamma_{{A}{B}_1{B}_2...{B}_N}$ estimated from the modulation data and detection data, shown in Fig. \ref{EBSmall} (b).

The other Bobs are the legitimate parities, this means their behaviors are restricted, where all Bobs are asked to perform coherent detection independently, and process the detection data only with Alice. Thus, the correlation  between Bob$_N$ and Bob$_j$ is $I({B}_N:{B}_j)$. What's more, this makes the other Bobs and Eve independent adversaries, which significantly reduces the cost of removing their knowledge on Alice and Bob$_N$.
Specifically, we can get the secret key rate for Bob$_N$ relative to Bob $j$ or Eve,
\begin{equation}
    \left \{
    \begin{aligned}
      & K_{({B}_N,{B}_j)} = H({B}_N)-I({B}_N:{B}_j)-leak_N, \\
      & K_{({B}_N,{E})} = H({B}_N)-\chi_{{B}_N{E}}-leak_N.
    \end{aligned}
    \right .
\end{equation}
Here, $H(X)$ is the Shannon entropy, and $leak_N$ is the information leakage during error correction.
Since the privacy amplification is processed according to the amount of the information known by the adversary, when the $N$ adversaries are independent, there is no need to remove the sum of the adversaries' knowledge, in contrast, removing the strongest adversary's knowledge is enough to make all adversaries unknown to the final secret key.
Therefore, the secret key rate between Alice and Bob$_N$ is
\begin{equation}
    \label{eq-K_N}
    \begin{aligned}
      K_{N} &= \min\left\{K_{({B}_N,{B}_1)}, ~K_{({B}_N,{B}_2)},~...~,~K_{({B}_N,{B}_{N-1})},\right. \\
      &\left. ~~~~~K_{({B}_N,{E})}\right\}, \\
      &=H({B}_N)-\max{\left\{\max_i{I({B}_N:{B}_i)},\chi_{{B}_N{E}}\right\}}-leak_N.
    \end{aligned}
\end{equation}
In Eq. \ref{eq-K_N}, Eve's knowledge as well as Bobs' correlations are removed simultaneously.

In error correction, the OTP encryption of the syndromes causes the loss of secret key. It is the same as the length of the syndromes $L_s=leak_N$.
Since the syndromes are encrypted, without the information leakage, the secret key actually generated is $K_N+L_s$, but the secret key with the length of $L_s$ is consumed for encrypting the syndrome. Thus, we can get the reconciliation efficiency with encrypted transmission of syndromes
\begin{equation}
    \label{beta}
    \beta = \frac{H({B}_N)-L_s}{I({A}:{B}_N)}.
\end{equation}
Note that, $\beta$ is the same as that when the syndrome is publicly transmitted in a one-way protocol. It provides an efficient way to perform error correction without extra leakage of Bobs' correlation.

In particular, when $\max\limits_i{I({B}_N:{B}_i)}<\chi_{{B}_N{E}}$, the removal of the correlation between Bobs can be included in the removal of the upper bound of the eavesdropped information. Thus, the secret key rate of Bob$_N$ has the same form as the traditional one \cite{DWBound}, with $\chi_{{B}_N{E}}$ estimated by more trusted modes,
\begin{equation}
    \label{eq2}
    \begin{split}
        K_N= \beta I\left({A}:{B}_N\right)-\chi_{{B}_N{E}}.
    \end{split}
\end{equation}
This means the correlations between the receivers contributes to a tight parameter estimation and can be removed without extra resources, which is conductive to achieve  high key rate.

The security analysis mentioned above need to deal with the covariance matrix containing massive modes, for simplification, a general method is proposed to reduce the $N+1$ mode matrix to a 3 mode one, detailed in Supplementary Note 1. In particular, for the channel structure shown in Fig. \ref{Structure}, we can get an EB scheme with just two outputs following the simplification method, shown in Fig. \ref{EBSmall} (c). The information in modes $\hat{B}_1$, $\hat{B}_2$, ..., $\hat{B}_{N-1}$ are concentrated into one mode, $\hat{B}_R$, the variance of $\hat{B}_R$ and its covariance with the other modes is enhanced, reflected in Fig. \ref{EBSmall} (d).

\begin{figure*}[t]
    \includegraphics[width=18 cm]{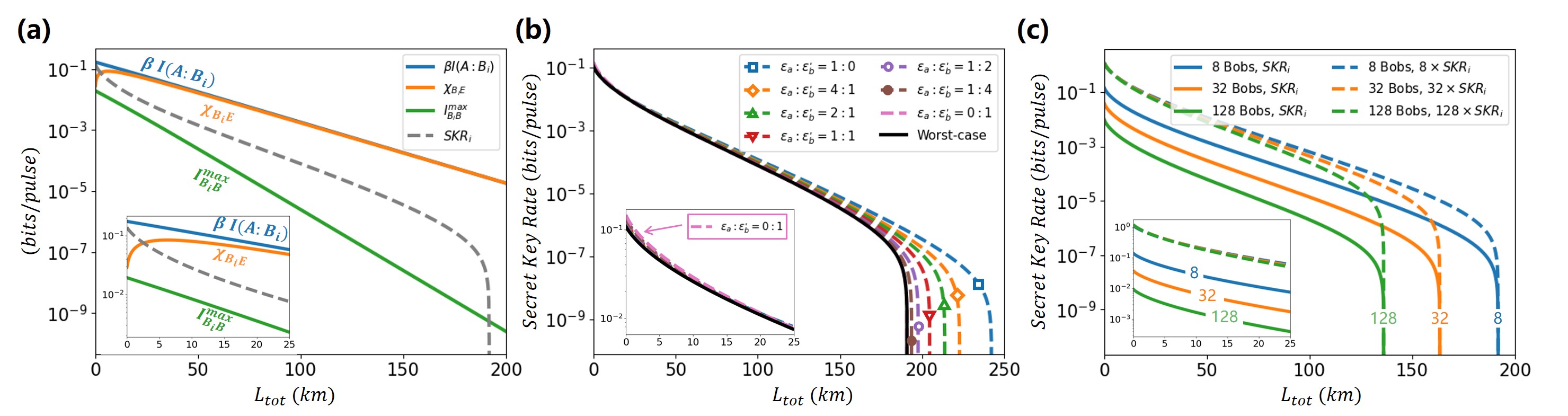}
    \caption{\label{Simulation} \textbf{Performance of the PTMP protocol.} Simulation result in symmetry case, where the length and the amount of the excess noise of the quantum channels connecting the power splitter with Bobs ($Ch_{b_i},i=1,2,...,N$) are the same. (a) Comparison between $I({A}:{B}_i)$ (blue line), $\chi_{{B}_i{E}}$ (orange line), $I^{max}_{{B}_i{B}}$ (green line), and the secret key rate of Bob $i$ ($SKR_i$, gray dashed line) with 8 receivers. (b) Comparison of secret key rate of Bob $i$ when $\varepsilon_{a}$ and $\varepsilon_{b}'$ have different ratio (dashed line), while $\varepsilon_{tot}$ is fixed, there are 8 receivers, and the worst-case secret key rate is presented (solid line). (c) Comparison of the secret key rate when the number of the Bobs is 8 (blue), 32 (orange) and 128 (green), the secret key rate of Bob $i$ ($SKR_i$, solid line) and the  secret key rate of the overall protocol (dashed line) are presented.
    The simulation parameters are as below, modulation variance $V_{M}=4$, reconciliation efficiency $\beta=95.6 \%$, $\varepsilon_{tot}=\varepsilon_{a}+\varepsilon_{b}'=0.0383$, $\varepsilon_{a}=0.004$ (for (a) and (c)), the detection efficiency $\eta_d=60 \%$, the detection noise $v_{el}=0.15$ \cite{CvExp80km2012,CvExp202kmPRL}. The excess noise of $Ch_{b_i}$ are the same, which can be calculated with $\varepsilon_{b}'T_1/N$.}
\end{figure*}

The simplification is based on the combination of two output modes, for a special case as in Supplementary Note 2,  where two modes come from the outputs of a beam splitter with transmittance $\eta$ and respectively experienced two independent channels with the transmittance and excess noise of $(T,\varepsilon)$ and $(T',\varepsilon')$, we can concentrate them to one mode which experienced a channel with
\begin{equation}
    \left\{
    \begin{aligned}
        T_t&=\eta T+(1-\eta)T',\\
        \varepsilon_t&=\frac{\eta T^2\varepsilon+(1-\eta)T'^2\varepsilon'}{(\eta T+(1-\eta)T')^2}.
    \end{aligned}
    \right.
\end{equation}
In particular, for a symmetrical case where $\eta=0.5$, $T=T'$ and $\varepsilon=\varepsilon'$, we can get $T_t=T=T'$, $\varepsilon_t=\varepsilon=\varepsilon'$. By paring the output modes, we can reduce the output modes easily.
Moreover, the simplified security analysis in this symmetrical case will not cause the loss of secret key rate, which can help to make the simulation easily for evaluating the performance of the protocol. The detailed derivation of the simplified EB scheme can be found in Supplementary Note 2, and the corresponding security analysis is shown in Supplementary Note 3.

~\\
\textbf{Performance of the protocol.}
~\\
Here, we present the channel situation which is consistent with the actual implementation scenario, with a structure  shown in Fig. \ref{Structure}, Eve attacks all of the quantum channels, $Ch_a, Ch_{b_1}, Ch_{b_2}, ...,Ch_{b_N}$, but introduces no correlations between $Ch_{b_i}$s.
Note that, if Eve really introduces correlations between $Ch_{b_i}$, the correlations can still be reflected in $\gamma_{AB_1B_2\ldots B_N}$, the security analysis strategy for the proposed protocol based on covariance matrix is general.

For simplification, in the simulations, the parameter of each $Ch_{b_i}$ are the same, which is a symmetry case. We define $\varepsilon_{tot}=\varepsilon_{a}+\varepsilon_{b}'$, with $\varepsilon_{b}=\varepsilon_{b}'T_1/N$, where $\varepsilon_{a}$ and $\varepsilon_{b}$ are the excess noise introduced in $Ch_a$ and $Ch_{b_i}$, $T_1$ is the transmittance of $Ch_a$.
Here, $\varepsilon_{a}$ corresponds to the noise which has the common effect on all of the Bobs, such as the noise due to modulation and the background noise. $\varepsilon_{b}$ represents the noise has independent effect on Bobs, such as the phase recovery noise.
The rationality of the setting of the excess noise in the PTMP protocol is explained in Supplementary Note 4.

Fig. \ref{Simulation} (a) shows that, at all of the secure transmission distance, $\chi_{B_iE}>I^{max}_{B_iB}$, thus all of the Bobs can generate uncorrelated secret key with Alice with the same original signals without extra cost. This significantly weakens the impact of Bobs' correlation on the performance of the protocol.

In Fig. \ref{Simulation} (b), the influence of $\varepsilon_{a}$ and $\varepsilon_{b}$ is shown.
When $\varepsilon_{tot}$ is fixed and the proportion of $\varepsilon_{Ch_b}'$ is increasing, the secret key rate decreases seriously at long distances, but slightly increases at a short distance, and the performance at short distance is rather stable. Even in the worst case, the network with 8 users can still transmits over 180 km.

In Fig. \ref{Simulation} (c), as the number of Bobs increases, the maximal transmission distance and the secret key rate of a single Bob is reduced, but the overall secret key rate of the protocol, which is $N$ times that of a single Bob, is rather stable.
Within 50 km, the overall secret key rate is hardly affected by the increase of the Bobs, which makes it have great access capability.
For the maximum transmission distance, even for 128 Bobs, the distance can still exceed 120 km, where the loss of the link is more than 45 dB, including the loss caused by an ideal optical power splitter (21 dB), and a 120 km fiber (24 dB).

Aiming at the practical implementation, the PTMP protocol can be used as a quantum access network \cite{QNetNature2013,QAN2015,QCnetOE2021} with PON. Here, the most important issue which may affect the practical network, the failure of error correction, is analyzed.
As mentioned before, the syndromes ($S_N$) are encrypted with the accumulated secret key ($K_N'$) by one time pad. When the error correction is performed successfully, the amount of the secret key consumed for syndrome encryption is compensated. However, once it fails, no secret key is generated in this round, thus the secret key for encrypting the syndromes is wasted, which greatly reduces the secret key rate of the overall network.

To deal with this problem, a key recycling strategy is proposed.
In the error-correction-failure round, the secrecy of $K_N'$ is actually not fully disclosed, since it's masked by $S_N$. This makes $K_N'$ can be seen as a weak secret key partially known by Eve, where the mutual information between Eve and $K_N'$  is
$I(E:K_N')\leq \chi_{{B}_N{E}}$, detailed in Methods. After a proper privacy amplification, the wasted secret key bits can be recycled.
If the failure probability of the PTMP error correction between Bob$_N$ and Alice is $p_{f}$, Eq. \ref{eq2} can be corrected as
\begin{equation}
    \label{eq6}
    \begin{split}
        K_i=&(1-p_{f})[\beta I\left({A}:{B}_i\right)-\chi_{{B}_i{E}}]-p_{f} \ \chi_{{B}_i{E}}\\
        =&(1-p_{f})\beta I\left({A}:{B}_i\right)-\chi_{{B}_i{E}} \\
        =&\beta_t I\left({A}:{B}_i\right)-\chi_{{B}_i{E}}.
    \end{split}
\end{equation}
Here, $\beta_{t}=(1-p_{f})\beta$. The failure probability of the error correction is reflected to the loss of reconciliation efficiency.
With the state of the art error correcting code, the optimal value of $\beta_t$ is about 90\% \cite{ErrorCorrection}.

\begin{figure}
    \includegraphics[width=8 cm]{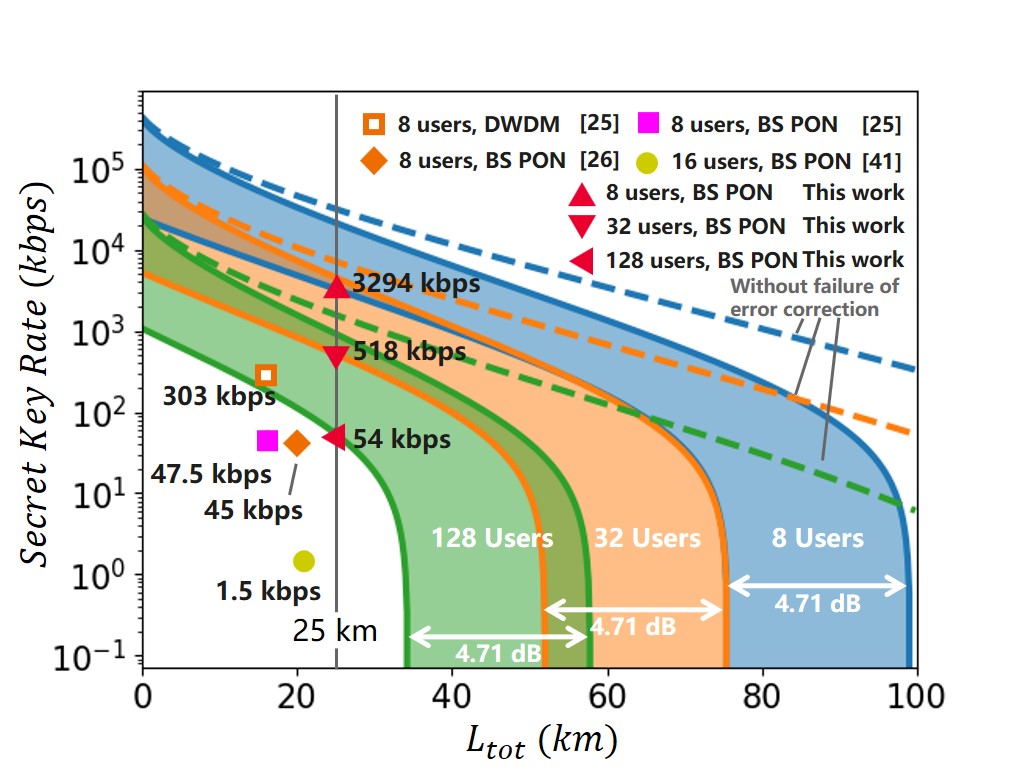}
    \caption{\label{PracNet}
    \textbf{The structure and secret key rate of the quantum secure access network.} The simulation results of the secret key rate of a single user. The colored area represents the fluctuation of the performance. Here, the maximal extra loss is settled as 4.71 dB, the modulation variance is optimized as 2.52, the repetition rate is settled as 5 GHz. The other simulation parameters are the same as before.}
\end{figure}

A simulation reflecting the practical performance of the quantum access network, including the error-correction-failure induced reconciliation efficiency reduction, is presented in Fig. \ref{PracNet}.
To reflect the extra loss of the devices in practical case, such as the imperfect connectors of the fibers and the extra loss of the optical power splitter, refer to \cite{QCnetOE2021}, an extra loss of $L_e=$ 4.71 dB which is assumed controlled by Eve is added to the analysis.
The other simulation parameters are all from the existing experiment of the practical implementation of CV-QKD \cite{CvExp5GRep,CvExp202kmPRL}.
Even in the worst case of this situation, the network can still access 128 users with a transmission distance longer than 25 km, which is a typical distance of an access network, and can achieve the secret key rate of 54 kbps per user at 25 km.
With fewer access points such as 32 and 8 users, since the loss caused by the optical power splitter is reduced, secret key rate in the worst case is 518 kbps and 3294 kbps per user under 25 km. This is far more than the existing quantum access network experiments with similar network structure, single photon detection and two-user protocols \cite{QNetNature2013,QAN2015,QCnetOE2021}, showing the potential key rate improvement of two orders of magnitude when supporting 8 users.

This enhancement is achieved due to two factors. The first is the tight etimation of the channel situation by using all receivers' detection data. It can use more receiver modes for security analysis than the two-user protocol, helping to reduce the effect of the loss from the optical power splitter. The second is the simultaneously secret key generation between the transmitter and all receivers, where each quantum signal with continuous-variable information can support multiple users, the availability is much higher than the qubit-implemented protocols.

~\\
\textbf{Discussion}
~\\
In this work, we have proposed a PTMP QKD protocol to simultaneously support multiple independent QKD links with one source. Now, QKD can support multiple users natively.
In the proposed protocol, each signal proposed by the transmitter can be used by all receivers to generate independent secret keys. This provides a solution in QKD protocol layer for interconnecting multiple users securely.
The proposed protocol makes full use of the information from the source, which contributes to a tight estimation of the channel and high key rate for both a single user and the overall network.

Compared to the current metropolitan networks with point-to-point connections, where 49.5 kbps key rate at 18 km \cite{46node}, and 49.4 kbps with 2.1 dB loss \cite{QCnetNature2021} is achieved, our protocol can support 128 users with 145 kbps for each user at 18 km, considering the failure of error correction and the extra loss of the fibre links.
Even if the repetition rate is reduced by 100 times, a 50 MHz transmitter can also support 8 users with 56.6 kbps of key rate for each user. Therefore, compared with the existing point-to-point protocols, our protocol is more suitable for quantum secure networks.

\begin{figure}
    \includegraphics[width=8 cm]{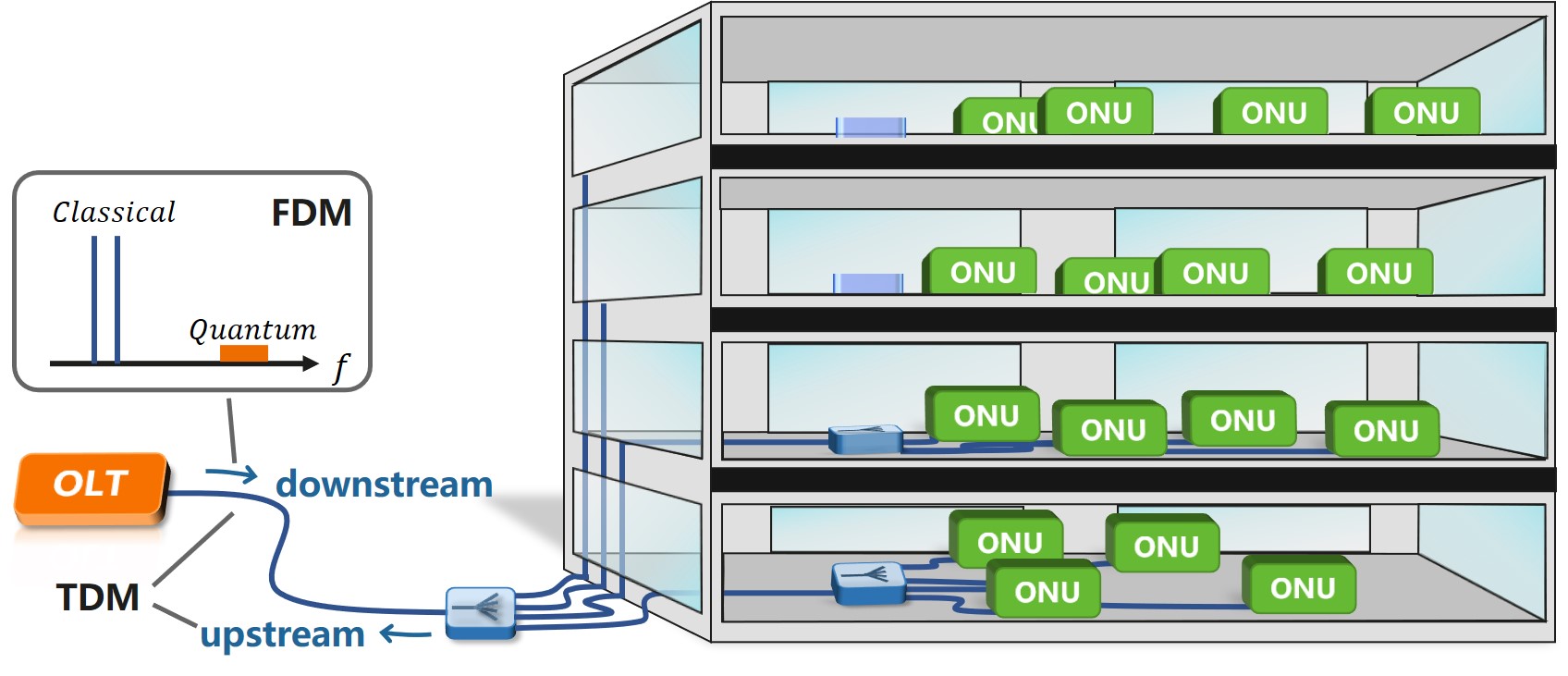}
    \caption{\label{Discuss}
    \textbf{The application scenario of the PTMP protocol using as a quantum access network.} The optical line terminal (OLT) produces classical and quantum signals with FDM. The optical network units (ONUs) receive and demodulate the signals, extracting classical signals and quantum keys. There are also upstream signals coexist in the network with TDM.}
\end{figure}

For the large-scale practical deployments in access and metropolitan distances, the PTMP protocol can be easily realized with a downstream PON, which is widely used in the classical optical network, as shown in Fig. \ref{Discuss}.
Since coherent detection can satisfy the decoding of quantum signals, by integrating the protocol in classical networks, the costs of the massive deployment of end users can be significantly reduced.
Moreover, because our protocol can support multiple users without multiplexing technologies, the frequency division multiplexing or wavelength division multiplexing can be used to get higher key rate rather than supporting multiple QKD links. These advantages result in a high-rate and cost-effective quantum secure network.

~\\
\textbf{Acknowledgments.}
~\\
This research was supported by the National Natural Science Foundation of China under Grants 62001044 and 62201013, the Equipment Advance Research Field Foundation under Grants 315067206, and the Fund of State Key Laboratory of Information Photonics and Optical Communications under Grants IPOC2021ZT02.




~\\
 \textbf{Methods}
 ~\\
 \textbf{Recycling the wasted secret key when error correction failed.}
 ~\\
 When the error correction is failed, using the secret key bit string $K'$ to encrypt the syndrome $S$ with OTP is also a process where the secret information of $K'$ is partially covered by $S$.
 In this process, no secret information is further leaked because no secret key is generated, the system has no output.
 Therefore, if we can get the lower bound of the secret information of $K'$, with privacy amplification, a secure key bit string can be regenerated, reducing the wasted secret key bits due to the failure of error correction.

 Since $K'$ is the quantum secret key, each bit is secure, random and independent, which cannot be accessed by Eve. Assuming Eve can get $S\oplus K'$, her knowledge about $K'$ is upper bounded by her knowledge on $S$.
 Take the LDPC code as an example, the syndrome $S=Hd$, here $d$ is the data after quantification with length $n$, $H$ is the parity-check matrix. Normally, $H=(A,I_{n-k})$, where $A$ is an $(n-k)\times k$ binary matrix, and $I_{n-k}$ is the identity matrix. By written $d$ as $(d_1, d_2)^T$, $S=Ad_1\oplus I_{n-k}d_2$, here $\oplus$ means Exclusive OR. With $S_1 = Ad_1$, $S_2=I_{n-k}d_2$, $S$ can be written as $ S_1 \oplus S_2$.
 Thus, Eve's knowledge on $S$ depends on her knowledge on $S_1$ and $S_2$.

 Since $S_2=I_{n-k}d_2=d_2$, Eve's knowledge on $S_2$ is the same as that on Bob's detection data, which can be upper bounded by the Holevo bound  $\chi_{BE}$.
 In the worst case, assuming that Eve has full knowledge on $S_1$, then Eve's knowledge on $S_1 \oplus S_2$ is the same as that on $S_2$, explained as below.

 Assuming two random variables with binomial distribution, $A$ and $B$, and Eve has full knowledge on $A$. Therefore, for the  random variable $C = A \oplus B$, Eve can establish a correspondence between $C$ and $B$, resulting in the mutual information $H(E:C)=H(E:B)$.
 When the random variables $A$ and $B$ correspond to $S_1$ and $S_2$ respectively, we can get the above conclusion.
Therefore, $I(E:S_1\oplus S_2)\leq \chi_{BE}$. Finally, we can get $I(E:K')\leq \chi_{BE}$.

~\\
\textbf{Accumulating the secret key in advance.}
~\\
Since the error correction requires encryption of the syndrome, the secret key bits should be accumulated before the protocol to support the first round of post-processing. It can be stored in the devices before deploying the network, or in the facilities such as secret key pool, where the legitimate parties have access to it.

Besides the methods mentioned above, Alice and each Bob can distribute secret key in a one-way like process through the network, providing a more efficient way to accumulate the secret key.
They still use the network to transmit the quantum states, but different like the PTMP protocol, only one Bob uses his data to generate secret key with Alice rather than all Bobs. Thus, in post-processing, only one syndrome from one Bob is generated, Alice and Bob can perform the post-processing like a one-way protocol, without encrypting the syndrome. Therefore, they don't have to worry about the effect due to error-correction-failure.
The security analysis still considers the whole network, thus the secret key rate of a single users is not affected.


\bibliography{apssamp}

\clearpage

\begin{widetext} 
    \section*{Supplementary Information}
    In this supplementary document, the simplification process of the network and the security analysis are detailed, including the general simplification method to reduce the modes in covariance matrix for a simpler security analysis, an equivalent simplified entanglement based scheme, and the security analysis of  the protocol.
    The rationality of the simulation parameters in the results is also discussed. 
    \section{Supplementary Note 1: General simplification method for covariance matrix}
    
     The main idea in security analysis is to construct a covariance matrix which contains all of the Bobs' modes and Alice's mode. However, with the increase of the number of Bobs, the modes of the matrix increases, and the secret key rate calculation may become complex. 
     To simplify the security analysis, we propose a method to decrease the modes of the matrix.
    
     Take the secret key rate calculation of Bob $N$ as an example.
     When calculating the secret key rate of Bob $N$, we focus on the mode $\hat{B}_N$, and we want to reduce the other modes to one mode to get a smaller matrix with only 3 modes. What's more, we hope the loss of the secret key rate due to the simplification should be small as well, thus the simplification of the security analysis will not affect the secret key rate too much.
    
     \begin{figure*}[b]
        \includegraphics[width=14 cm]{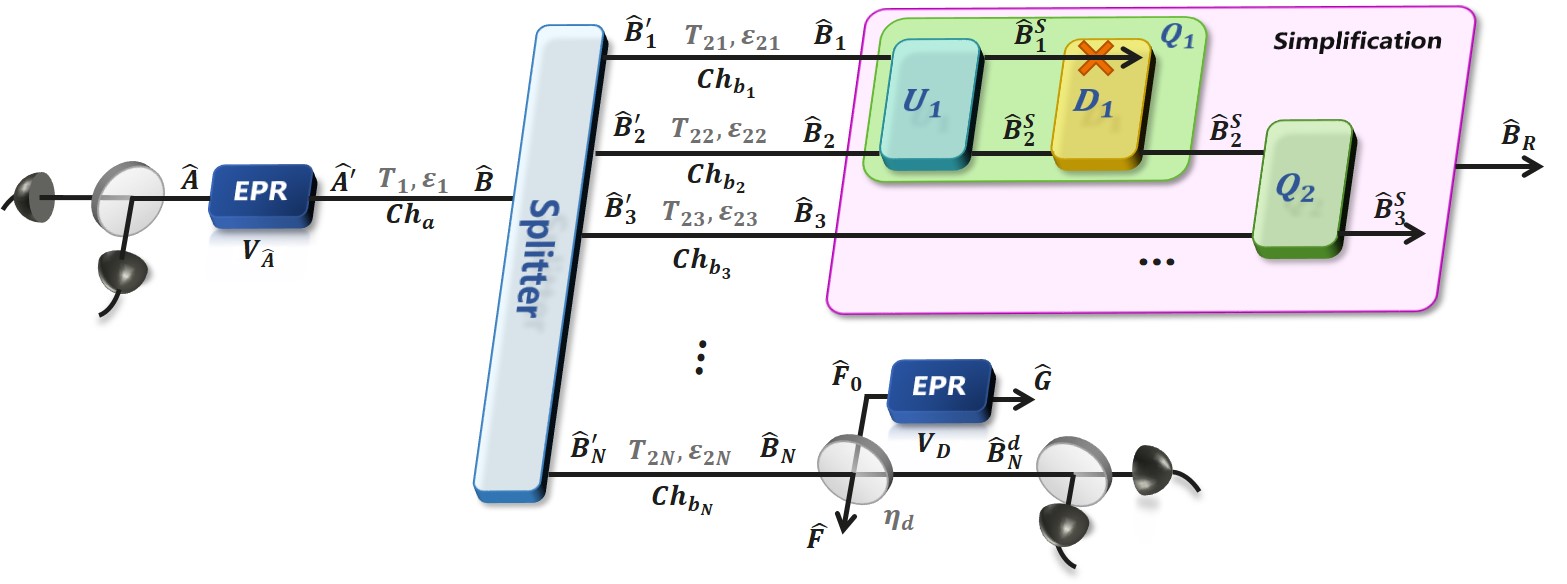}
        \caption{\label{EBScheme}
        \textbf{The entanglement-based (EB) scheme representing the simplification process.} Alice prepares EPR state with variance $V_{{A}}$,  mode $\hat{A}$ is heterodyne detected, and the other mode $\hat{A'}$ is sent to the quantum channel $Ch_a$ with transmittance $T_1$ and excess noise $\varepsilon_1$. After passing through $Ch_a$, mode $\hat{A'}$ transforms into mode $\hat{B}$, and then divided by the splitter into $N$ modes, $\hat{B_1'}$, $\hat{B_2'}$, $...$, $\hat{B_N'}$. The modes then pass through $Ch_{b_1}$, $Ch_{b_2}$, $...$, $Ch_{b_N}$, and finally detected by different Bobs. 
        The simplification process consists of the simplification unit for 2 modes, $Q$, including a unitary transformation $U_1$, and the removal operation of one mode, written as $D_1$.
        After repeating the operations above, the other modes $\hat{B}_1, \hat{B}_2, ..., \hat{B}_{N-1}$ can be combined as one mode, $\hat{B}_R$.
        }
    \end{figure*}
    
     Directly reducing the modes will cause a huge loss of secret key rate, thus we first transform and recombine the modes, then cut the other output modes.
     As shown in Fig. \ref{EBScheme}, it's possible to construct a unitary transform $U_1$ which acts on two of the other $N-1$ receiver modes (e.g. $\hat{B}_{1}$, $\hat{B}_2$) to decrease the correlation between one output mode ($\hat{B}_1^S$) and mode $\hat{A}$, while increasing the correlation between the other output mode ($\hat{B}_{2}^S$) and mode $\hat{A}$ as below
     \begin{equation}
         \left\{
         \begin{aligned}
             \gamma_{{{A}{B}}_N {B}_{N-1} \ldots {B}_{2}^S{B}_1^S}&=U_1 \gamma_{{{A}{B}}_N{B}_{N-1} \ldots {B}_{2}{B}_1}U_1^T\\
             C_{{A}{B}_{1}^S}&<C_{{A}{B}_1}\\
             C_{{A}{B}_{2}^S}&>C_{{A}{B}_{2}}
         \end{aligned}
         \right.
         ,
     \end{equation}
     where $C_{{X}{Y}}$ represent the correlation between two modes $(\hat{X}, \hat{Y})$.
     The secret key rate calculated by $\gamma_{{{A}{B}}_N {B}_{N-1} \ldots {B}_{2}^S{B}_1^S}$ and $\gamma_{{{A}{B}}_N{B}_{N-1} \ldots {B}_{2}{B}_1}$ are the same, because $U_1$ is a unitary transform. However, if remove the mode $\hat{B}_1^S$,
     \begin{equation}
         \label{eqSimpl}
         \gamma_{{{A}{B}}_N {B}_{N-1} \ldots {B}_{2}^S} = D_{1} \ \gamma_{{{A}{B}}_N {B}_{N-1} \ldots {B}_{2}^S{B}_1^S} \ D_{1}^T,
     \end{equation}
     here $D_{1}$ is a $2N \times 2(N+1)$ matrix, with $D_{1}=(I_{2N},\ Z)$, $I_{2N}$ represents an $2N\times 2N$ identity matrix, and $Z$ represents a  $2N\times 2$ zero matrix. 
     The secret key rate calculated with $\gamma_{{{A}{B}}_N {B}_{N-1} \ldots {B}_{2}^S}$ becomes lower because the removal of the mode means that we don't trust it any more, and we loss the information.
     This process is detailed in Fig. \ref{CovMat}.  
    
     Define $D_{1}U_1$ as $Q_1$, thus, eq. \ref{eqSimpl} can be rewritten as
     \begin{equation}
        \gamma_{{{A}{B}}_N {B}_{N-1} \ldots {B}_{2}^S} = Q_1 \gamma_{{{A}{B}}_N{B}_{N-1} \ldots {B}_{2}{B}_1} Q_1^T
        .
     \end{equation}
     Further, by repeating this process, we can get
     \begin{equation}
         \label{Simplification}
         \gamma_{{A}{B}_N{B}_{N-1}^S} = (\prod \limits_{i=N-2}^{1} Q_i) \gamma_{{{A}{B}}_N{B}_{N-1} \ldots {B}_{2}{B}_1} (\prod \limits_{i=1}^{N-2} Q_i^T).
     \end{equation}
     A simplified matrix with only three modes is achieved. 
     If we denote Bob$_N$'s secret key rate after the operation of $Q_i$ as $SKR^{s_i}_N$, we can get $SKR^{s_{N-2}}_N \le SKR^{s_{N-3}}_N \le ... \le SKR^{s_1}_N \le SKR_N$.
     The secret key rate calculated with $ \gamma_{AB_NB_{N-1}^S}$ is no more than that calculated with $\gamma_{{AB}_1B_2\ldots B_N}$, thus the operation is always secure.
    
    \begin{figure*}
        \includegraphics[width=14 cm]{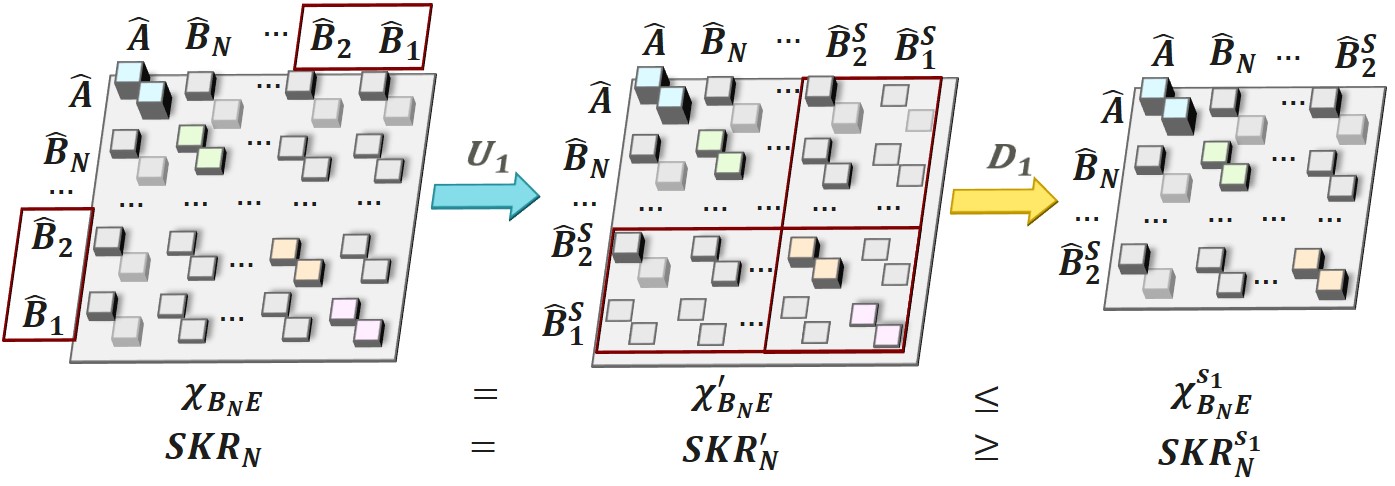}
        \caption{\label{CovMat}
        \textbf{The transformation of the modes in covariance matrix.} After a unitary transformation, the modes $\hat{B}_1$ and $\hat{B}_2$ are transformed to $\hat{B}_1^S$ and $\hat{B}_2^S$. The correlation between $\hat{B}_1^S$ and the other modes are reduced, which means $\hat{B}_1^S$ is decoupled. We then remove $\hat{B}_1^S$ and get a smaller covariance matrix. Here, the Holevo bound remains unchanged before and after $U_1$. After $D_1$, we loss some information about the channels, thus the estimation is not as tight as before, resulting in the increased Holevo bound and the decreased secret key rate of Bob$_N$ ($SKR_N$). 
        }
    \end{figure*}
    
    One of the $U_i$ is a transformation matrix of a beam splitter with transmittance $\eta_S$, written as 
    \begin{equation}
        Y_{BS} = 
        \begin{pmatrix}
            \sqrt{\eta_S} \cdot I_2 & -\sqrt{1-\eta_S} \cdot I_2\\
            \sqrt{1-\eta_S} \cdot I_2&\sqrt{\eta_S} \cdot I_2
           \end{pmatrix}
           .
    \end{equation}
    To simplify the analysis, we focus on the the modes $B_1$, $B_2$ and $A$, to explore the correlation between the receiver modes and the transmitter mode before and after the simplification.
    We define $\gamma_{{A}{B}_2{B}_1}$ as 
    \begin{equation}
        \gamma_{{A}{B}_2{B}_1} = 
        \begin{pmatrix}
            \gamma_{{A}}& C^T\\
            C &\gamma_{{B}_2{B}_1}
          \end{pmatrix}
          .
    \end{equation}
    Here, 
    \begin{equation}
        C=
        \begin{pmatrix}
            C_{{A}{B}_2}\sigma_z\\
            C_{{A}{B}_1}\sigma_z
           \end{pmatrix}
           ,
    \end{equation}
    \begin{equation}
        \sigma_z=
        \begin{pmatrix}
            1 & 0\\
            0 &-1
           \end{pmatrix}
           .
    \end{equation}
    The corresponding transformation $U_1$ is as below,
    \begin{equation}
        U_1=
        \begin{pmatrix}
            I_{2} & Z_2^T\\
            Z_2 &Y_{BS}
           \end{pmatrix}
           .
    \end{equation}
    Here, $Z_2$ represents the $4\times 2$ zero matrix.
    Therefore, we can get $\gamma_{{A}{B}_2^S{B}_1^S}=U_1 \gamma_{{A}{B}_2{B}_1} U_1^T$, which can be written as
    \begin{equation}
        \gamma_{{A}{B}_2^S{B}_1^S}=
        \begin{pmatrix}
            \gamma_{{A}}& C^T Y_{BS}^T \\
            Y_{BS}C & Y_{BS}\gamma_{{B}_2{B}_1}Y_{BS}^T
        \end{pmatrix}
          .
    \end{equation}
    Here, 
    \begin{equation}
        Y_{BS}C = 
        \begin{pmatrix}
            (\sqrt{\eta_S}C_{{A}{B}_2}-\sqrt{1-\eta_S}C_{{A}{B}_1})\sigma_z\\
           (\sqrt{\eta_S}C_{{A}{B}_1}+\sqrt{1-\eta_S}C_{{A}{B}_2})\sigma_z
           \end{pmatrix} .
    \end{equation}
    If set $\eta_S$ as 
    \begin{equation}
        \eta_S = C_{{A}{B}_1}^2/(C_{{A}{B}_1}^2+C_{{A}{B}_2}^2), (C_{{A}{B}_1} \cdot C_{{A}{B}_2} \ge 0),
    \end{equation}
    or 
    \begin{equation}
        \label{generalResults}
        \eta_S = C_{{A}{B}_2}^2/(C_{{A}{B}_1}^2+C_{{A}{B}_2}^2), (C_{{A}{B}_1} \cdot C_{{A}{B}_2} \le 0). 
    \end{equation}
    we can make $C_{{A}{B}_2^S}=0$ or $C_{{A}{B}_1^S}=0$. In this way, the correlation between one of the output modes and the transmitter mode is reduced to 0, which means the mode $\hat{B}_2^S$ can be seen as a noise mode introduced in receiver. Even if we ignore it during security analysis, the impact on secret key rate is not serious.

     In practical implementation, because the structure of the channel is not specified, thus the simplification units have to be calculated based on the matrix estimated, which is $Q_i$ in the general simplification strategy. Note that, as long as $Q_i=D_iU_i$, where $U_i$ is a unitary transformation and $D_i$ represents the decrease of the modes, the secret key rate calculated after simplification is lower than that calculated with the overall matrix, which means the simplification is always secure.
     The form of $U_i$ will affect the secret key rate calculated by the simplification unit, aiming at achieving high secret key rate, each $Q_i$ should be optimized. It will take much calculation for the first time, but for practical channels, since the fiber channels are usually stable, it is possible to design an algorithm to accelerate the optimization of $Q_i$ with the results in last round, which can promote the efficiency.
     
     To further simplify, one can construct a 2-user scheme with the worst channel parameters estimated with the matrix, or simply remove a few modes to reduce the calculation, though the performance may be reduced, the security analysis becomes rather easy.
     In simulations, we focus on the scene where the transmittance and the excess noise of the fibers is fixed, especially when all of the fibers have the same channel parameters.
     With the simplification method, we can get a simple entanglement-based scheme for simulation. What's more, when the channel parameters are the same, there is no loss of secret key rate when using the simplified method, which can help us to explore the performance of the protocol easily.
    
    \section{Supplementary Note 2: Simplified equivalent network scheme}
    In this part, we specify a case consistent with the actual implementation scenario, where Eve attacks all of the quantum channels, but introduces no correlations between them.
    \subsection{Mode transformation of the simplification unit}
    \begin{figure}
        \includegraphics[width=14 cm]{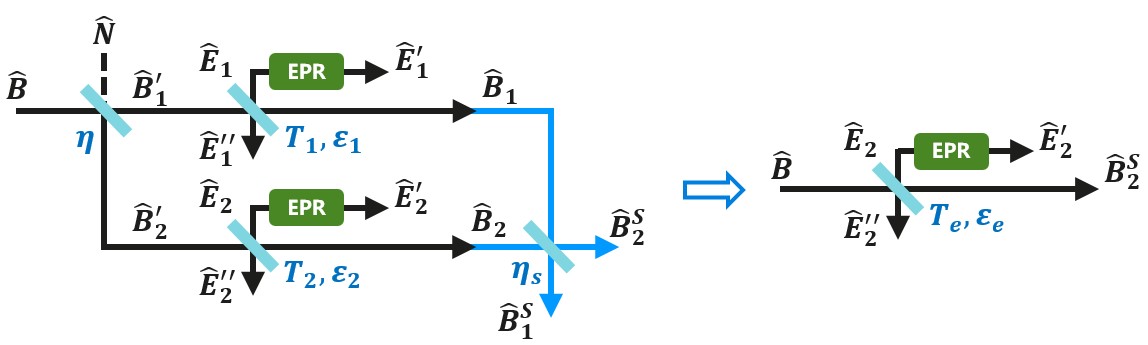}
        \caption{\label{SuppSimpUnit}
        \textbf{Combination of mode $\hat{B}_1$ and $\hat{B}_2$ with simplification unit.}}
    \end{figure}
    
    In this part, the derivation of the transform of the modes in the simplification unit is detailed. As shown in Fig. \ref{SuppSimpUnit}, the mode $\hat{B}$ is divided by a beam splitter with transmittance $\eta$, then the two output modes $\hat{B}_{10}$ and $\hat{B}_{20}$ are  sent into two different channels, with the channel parameters $(T_1, \varepsilon_1)$ and $(T_2, \varepsilon_2)$. This is the process the same as the quantum states division and transmission in the PTMP quantum secure network, mode $\hat{B}$ is the information mode, which has correlation with Alice's mode. $\hat{B}_1'$ and $\hat{B}_2'$ can be written as
    \begin{equation}
        \hat{B}_1'=\sqrt{\eta} \hat{B} +\sqrt{1-\eta}\hat{N},
    \end{equation}
    \begin{equation}
        \hat{B}_2'=\sqrt{\eta} \hat{N} -\sqrt{1-\eta}\hat{B}.
    \end{equation}
    Here, mode $\hat{N}$ is the vacuum state introduced by the beam splitter.
    After passing through the quantum channels, mode $\hat{B}_1'$ and mode $\hat{B}_2'$ are transformed to $\hat{B}_1$ and $\hat{B}_2$, the quantum channel is assumed fully controlled by eavesdropper Eve, but Eve doesn't introduce correlation between the two quantum channels. Assuming Eve's modes introduced to the quantum channels are $\hat{E}_1$ and $\hat{E}_2$, the mode $\hat{B}_1$ and $\hat{B}_2$ are detailed as below:
    \begin{equation}
        \hat{B}_1 = \sqrt{T_1}\hat{B}_1'+\sqrt{1-T_1}\hat{E}_1,
    \end{equation}
    \begin{equation}
        \hat{B}_2 = \sqrt{T_2}\hat{B}_2'+\sqrt{1-T_2}\hat{E}_2.
    \end{equation}
    Therefore, we can get the correlation between the transmitter mode and the receiver modes as 
    \begin{equation}
        C_{{A}{B}_1}=\sqrt{T_1}C_{{A}{B}_1'}=\sqrt{\eta T_1}C_{{A}{B}},
    \end{equation}
    \begin{equation}
        C_{{A}{B}_2}=\sqrt{T_2}C_{{A}{B}_2'}=-\sqrt{(1-\eta) T_2}C_{{A}{B}}.
    \end{equation}
    Based on Eq. \ref{generalResults}, we can get
    \begin{equation}
        \label{result_eta_S}
        \eta_S = (1-\eta)T_2/(\eta T_1+(1-\eta)T_2)
        .
    \end{equation}
    
    By acting on $\hat{B}_1$ and $\hat{B}_2$ a unitary transformation, which is the beam splitter with transmittance $\eta_S$, the mode $\hat{B}_1$ and $\hat{B}_2$ are transformed to $\hat{B}_{1}^S$ and $\hat{B}_{2}^S$. Thus, 
    \begin{equation}
        \hat{B}_{1}^S=\sqrt{\eta_S}\hat{B}_1+\sqrt{1-\eta_S}\hat{B}_2,
    \end{equation}
    \begin{equation}
        \hat{B}_{2}^S=\sqrt{\eta_S}\hat{B}_2-\sqrt{1-\eta_S}\hat{B}_1.
    \end{equation}
    
    With the equations above, $\hat{B}_1^S$ is detailed as:
    \begin{equation}
        \begin{split}
        \hat{B}_1^S =& \sqrt{\eta_S}\left(\sqrt{T_1}\hat{B}_1'+\sqrt{1-T_1}\hat{E}_1\right)+\sqrt{1-\eta_S}\left(\sqrt{T_2}\hat{B}_2'+\sqrt{1-T_2}\hat{E}_2\right) \\
        =& \sqrt{\eta_S}\left(\sqrt{T_1}\left(\sqrt\eta \hat{B}+\sqrt{1-\eta}\hat{N}\right)+\sqrt{1-T_1}\hat{E}_1\right)+\sqrt{1-\eta_S}\left(\sqrt{T_2}\left(\sqrt\eta \hat{N}-\sqrt{1-\eta}\hat{B}\right)+\sqrt{1-T_2}\hat{E}_2\right) \\
        =& \sqrt{\eta_ST_1\eta}\hat{B}+\sqrt{\eta_ST_1\left(1-\eta\right)}\hat{N}+\sqrt{\eta_S\left(1-T_1\right)}\hat{E}_1+\sqrt{\left(1-\eta_S\right)T_2\eta}\hat{N}-\sqrt{\left(1-\eta_S\right)T_2\left(1-\eta\right)}\hat{B}\\
        &+\sqrt{\left(1-\eta_S\right)\left(1-T_2\right)}\hat{E}_2\\
        =& \left(\sqrt{\eta_ST_1\eta}-\sqrt{\left(1-\eta_S\right)T_2\left(1-\eta\right)}\right)\hat{B}+\sqrt{\eta_S\left(1-T_1\right)}\hat{E}_1+\sqrt{\left(1-\eta_S\right)\left(1-T_2\right)}\hat{E}_2\\
        &+\left(\sqrt{\eta_ST_1\left(1-\eta\right)}+\sqrt{\left(1-\eta_S\right)T_2\eta}\right)\hat{N}.
        \end{split}
    \end{equation}
    By adjusting $\eta_S$ as Eq. \ref{result_eta_S},
    we can reduce the coefficient of mode $\hat{B}$ in mode $\hat{B}_1^S$ to 0
    \begin{equation}
        \sqrt{\eta_ST_1\eta}-\sqrt{\left(1-\eta_S\right)T_2\left(1-\eta\right)}=\sqrt{\frac{\left(1-\eta\right)T_2T_1\eta}{\eta T_1+\left(1-\eta\right)T_2}}-\sqrt{\frac{\eta T_1T_2\left(1-\eta\right)}{\eta T_1+\left(1-\eta\right)T_2}}=0,
    \end{equation}
    which is consistent with the previous conclusions.
    
    With the methods above, mode $\hat{B}$ in mode $\hat{B}_1^S$ is removed, thus
    \begin{equation}
        \hat{B}_1^S=\sqrt{\frac{\left(1-\eta\right)\left(1-T_1\right)T_2}{\eta T_1+\left(1-\eta\right)T_2}}\hat{E}_1+\sqrt{\frac{\eta T_1\left(1-T_2\right)}{\eta T_1+\left(1-\eta\right)T_2}}\hat{E}_2+\sqrt{\frac{T_1T_2}{\eta T_1+\left(1-\eta\right)T_2}}\hat{N}.
    \end{equation}
    $\hat{B}_2^S$ is written as
    \begin{equation}
        \label{GeneralB2S}
        \begin{split}
            \hat{B}_2^S=&-\left(\sqrt{\frac{\left(1-\eta\right)T_2}{\eta T_1+\left(1-\eta\right)T_2}T_2\left(1-\eta\right)}+\sqrt{\frac{\eta T_1}{\eta T_1+\left(1-\eta\right)T_2}T_1\eta}\right)\hat{B}\\
            &+\left(\sqrt{\frac{\left(1-\eta\right)T_2}{\eta T_1+\left(1-\eta\right)T_2}T_2\eta}-\sqrt{\frac{\eta T_1}{\eta T_1+\left(1-\eta\right)T_2}T_1\left(1-\eta\right)}\ \right)\hat{N}-\sqrt{\frac{\eta T_1}{\eta T_1+(1-\eta)T_2}(1-T_1)}\hat{E}_1\\
            &+\sqrt{\frac{(1-\eta) T_2}{\eta T_1+(1-\eta)T_2}(1-T_2)}\hat{E}_2\\
            =&-\frac{T_2\left(1-\eta\right)+T_1\eta}{\sqrt{\eta T_1+\left(1-\eta\right)T_2}}\hat{B}-\sqrt{\frac{\eta T_1}{\eta T_1+(1-\eta)T_2}(1-T_1)}\hat{E}_1+\sqrt{\frac{(1-\eta) T_2}{\eta T_1+(1-\eta)T_2}(1-T_2)}\hat{E}_2\\
            &+\sqrt{\frac{(1-\eta)\eta}{\eta T_1+(1-\eta)T_2}}(T_2-T_1)\hat{N}\\
            =&-\sqrt{\eta T_1+\left(1-\eta\right)T_2}\hat{B}-\sqrt{\frac{\eta T_1}{\eta T_1+(1-\eta)T_2}(1-T_1)}\hat{E}_1+\sqrt{\frac{(1-\eta)T_2}{\eta T_1+(1-\eta)T_2}(1-T_2)}\hat{E}_2\\
            &+\sqrt{\frac{(1-\eta)\eta}{\eta T_1+(1-\eta)T_2}}(T_2-T_1)\hat{N}.
        \end{split}
    \end{equation}
    It's obviously that the output mode $\hat{B}_1^S$ contains only the noisy modes, which are $\hat{E}_1$, $\hat{E}_2$, and $\hat{N}$, these modes has no correlation with the information mode $\hat{B}$.

    \subsection{Equivalent one-way channel parameters of the simplification unit}
    After getting mode $\hat{B}_1^S$ and $\hat{B}_2^S$, mode $\hat{B}_1^S$ which only contains the noisy modes should be removed to reduce the number of the modes in the network scheme. Without mode $\hat{B}_1^S$, only mode $\hat{B}_2^S$ is left, thus a virtual one-way channel can be built, reflecting the mode transformation from mode $\hat{B}$ to mode $\hat{B}_2^S$. Based on Eq. \ref{GeneralB2S}, the variance of $\hat{B}_2^S$ is 
    \begin{equation}
        \begin{split}
            V_{{B}_{2}^S}=&\left(\eta T_1+\left(1-\eta\right)T_2\right)V_{{B}}+\frac{\eta T_1}{\eta T_1 + (1-\eta)T_2}(1-T_1+T_1\varepsilon_1)+\frac{(1-\eta) T_2 }{\eta T_1+ (1-\eta)T_2}(1-T_2+T_2 \varepsilon_2)\\
            &+\frac{(1-\eta)\eta(T_2-T_1)^2}{\eta T_1+(1-\eta)T_2}.
        \end{split}
    \end{equation}
    Making $T_e = \eta T_1+(1-\eta)T_2$, $V_{{B}_{2}^S}$ can be written as 
    \begin{equation}
        \label{VBS2}
        \begin{split}
            V_{{B}_{2}^S}
            &=T_eV_{{B}}+1+\frac{1}{T_e}((1-\eta)\eta(T_2-T_1)^2-\eta T_1^2+\eta T_1^2\varepsilon_1-(1-\eta)T_2^2+(1-\eta)T_2^2\varepsilon_2).
        \end{split}
    \end{equation}
    Noticed that 
    \begin{equation}
       \begin{split}
        (1-\eta)\eta(T_2-T_1)^2-\eta T_1^2-(1-\eta)T_2^2 
       =-T_e^2.
       \end{split}
    \end{equation}
    Thus, Eq. \ref{VBS2} can be written as 
    \begin{equation}
        V_{\hat{B}_{2}^S}=T_eV_{\hat{B}}+1+\frac{1}{T_e}(-T_e^2+\eta T_1^2\varepsilon_1+(1-\eta)T_2^2\varepsilon_2).
    \end{equation}

    \begin{figure}
        \includegraphics[width=18 cm]{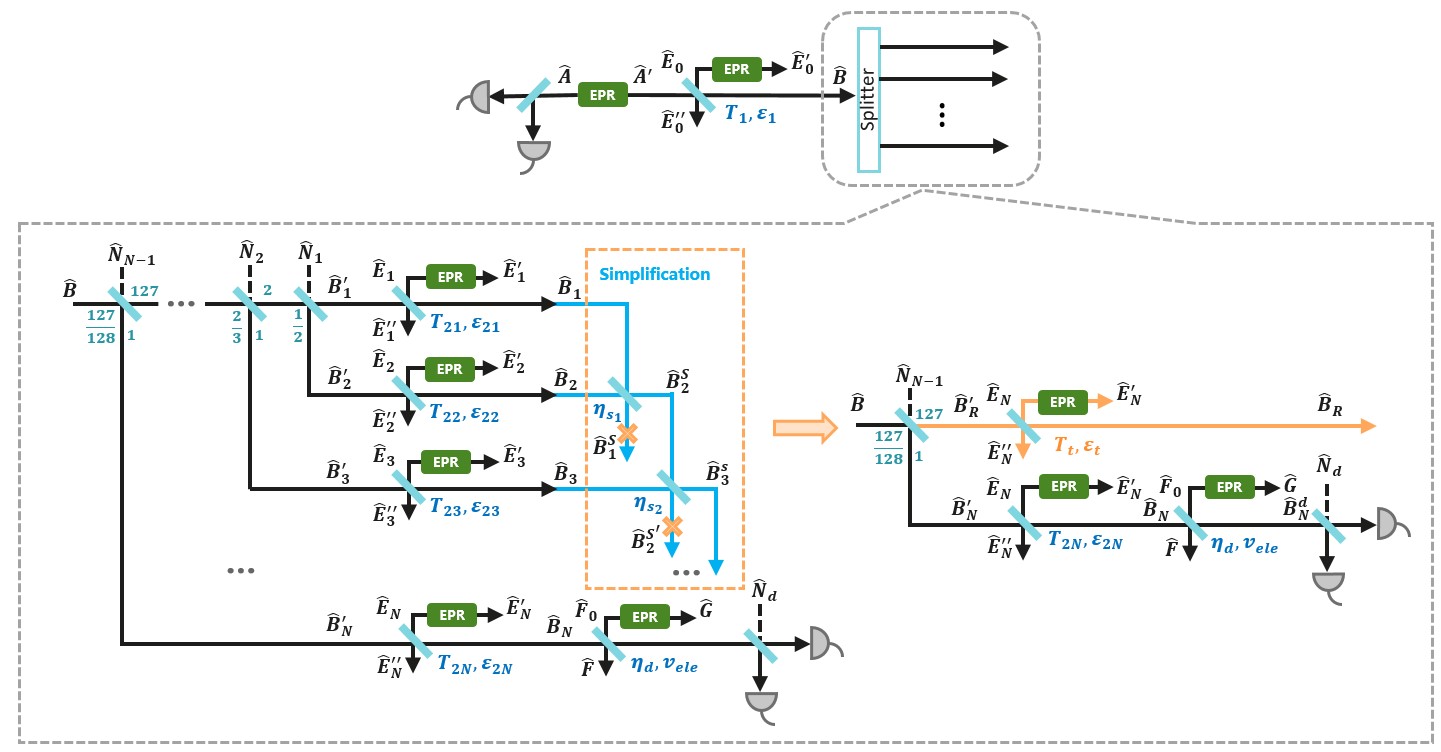}
        \caption{\label{SuppSimpEB}
        \textbf{Simplifying the network scheme to a 2-user scheme with the simplification unit.} The PTMP network consists of a passive optical network with optical power splitter. One of the two modes from the EPR state produced by Alice is heterodyne detected, while the other one is sent to the quantum channel with $T_1, \varepsilon_1$. The attack of the eavesdropper Eve is modeled by an  EPR state with beam splitter. She performs one-mode attack, which introduces no correlations between different channels of the network. The output of the splitter is connected with different quantum channels respectively, which is attacked by Eve. These channels have the parameters written as $T_{2i}, \varepsilon_{2i}$. For an N-user network, there are $N$ receiver modes in the EB scheme, which are $\hat{B}_1, \hat{B}_2, ..., \hat{B}_N$. With the simplification shown in the amber box, the receiver modes are recombined and removed. After simplification, the receiver modes except for which the secret key rate is calculated are transformed to one mode $B_R$. Here, assuming the secret key rate of mode $B_N$ is being calculated.}
    \end{figure}

    Finally, a standard form of $V_{\hat{B}_2^S}$ can be achieved
    \begin{equation}
        V_{\hat{B}_{2}^S}=T_eV_{\hat{B}}+1-T_e+T_e\left(\frac{\eta T_1^2\varepsilon_1+\left(1-\eta\right)T_2^2\varepsilon_2}{T_e^2}\right).
    \end{equation}
    This is the standard variance formula of a mode after passing through the one-way channel with transmittance
    \begin{equation}
        T_e = \eta T_1+(1-\eta)T_2,
    \end{equation}
    and excess noise
    \begin{equation}
        \varepsilon_e = \frac{\eta T_1^2\varepsilon_1+\left(1-\eta\right)T_2^2\varepsilon_2}{T_e^2}.
    \end{equation}
    These are the equivalent one-way channel parameters of the simplification unit in general. For $T_e$, this is the average of $T_1$ and $T_2$, affected by the uniformity of the beam splitter with transmittance $\eta$. For $\varepsilon_e$, this is the average of $\varepsilon_1$ and $\varepsilon_2$, affected by $\eta$ and $T_i^2$.
    Further, when $T_1=T_2=T$, based on the above equation, the equivalent parameters are as below
    \begin{equation}
        T_e=T,\ \varepsilon_e = \eta \varepsilon_1+(1-\eta)\varepsilon_2.
    \end{equation}
    If $T_1=T_2=T$ and $\varepsilon_1=\varepsilon_2=\varepsilon$, then 
    \begin{equation}
        T_e = T, \varepsilon_e = \varepsilon.
    \end{equation}
    It is obvious that when the transmittance ($T_1,\ T_2$) or the excess noise ($\varepsilon_1,\ \varepsilon_2$) of the two channels are equal, the equivalent one-way channel parameters are not affected by the transmittance of the beam splitter $\eta$.
    Thus, in the symmetrical case when the channels have the same parameters ($T$ and $\varepsilon$), the equivalent one-way channel parameters are $T$ and $\varepsilon$ as well, no matter whether $\eta=0.5$.
    
    \subsection{Simplification of the security analysis when the channels are not correlated}

    With the simplification unit as above, the entanglement-based (EB) scheme of the network can be simplified to a 2-user scheme as shown in Fig. \ref{SuppSimpEB}. The  EB scheme shown here specifies a network scheme for simulation. In practical implementation where the modulation data and the detection data can be used for security analysis, the specific scheme of the network is unnecessary.
    The network has $N$ receiver modes $\hat{B}_1, \hat{B}_2, ..., \hat{B}_N$, the key for simplification is to reduce the modes in the scheme thus to simplify the security analysis. The simplest way to remove the modes is directly reduce the modes in the matrix, but it will cause the serious decrease of the performance of the network because the removal of the modes means that the modes are untrusted.  A reasonable strategy is to recombine the receiver modes by unitary transformation before removing them. The unitary transformation will not affect the secret key rate, but it can change the relationship and correlation between different modes. After that, with a carefully designed unitary transformation, the removing of the modes after transformation will not cause much loss. Thus the simplification of security analysis with less impact on secret key rate is realized.
    
    In Fig. \ref{SuppSimpEB}, for simplification, mode $\hat{B}_1$ and $\hat{B}_2$ are first coupled with a beam splitter with transmittance $\eta_{S_1}$. The output modes are $\hat{B}_1^S$ and $\hat{B}_2^S$, by adjusting the transmittance of the beam splitter, $\eta_{S_1}$, as mentioned above, the correlation between mode $\hat{B}_1^S$ and the other modes can be reduced. Thus, removing the mode $\hat{B}_1^S$ will affect less on the secret key rate of the network. After removing mode $\hat{B}_1^S$, the equivalent one-way channel can be built, thus we can calculate a proper transmittance $\eta_{S_2}$ to transform mode $\hat{B}_{2}^S$ and $\hat{B}_3$ to mode $\hat{B}_2^{S'}$ and $\hat{B}_3^S$, where the correlation between $\hat{B}_2^{S'}$ and the other modes is reduced. Thus, mode $\hat{B}_2^{S'}$ can be removed without much impact on secret key rate. By repeating the process as above, the modes except for $\hat{B}_N$ are finally transformed to one mode $\hat{B}_R$, with the total equivalent one-way parameters $T_t$ and $\varepsilon_t$. When Eve performs one-mode attack, and  $T_2=T_{21}=T_{22}=...=T_{2N}$, $\varepsilon_{2}=\varepsilon_{21}=\varepsilon_{22}=...=\varepsilon_{2N}$, the equivalent one-way parameters of the 2-user scheme is the same as the channel parameters, $T_t=T_2$, $\varepsilon_{t}=\varepsilon_2$. What's more, the correlation between the modes to be removed, $\hat{B}_1^S$, $\hat{B}_2^{S'}$, et. al, is zero. Thus, the removal of the modes will no cause loss of security, a perfect simplification can be performed. 
    In conclusion, the simplification EB scheme can play a role in the simulation of the network, it significantly simplifies the numerical analysis, what's more, the simplification in some situations has no secret key rate loss, the performance of the network can be reflected accurately.

    \section{Supplementary Note 3: Detailed secret key rate calculation with the 2-user scheme}
    \begin{figure}
        \includegraphics[width=18 cm]{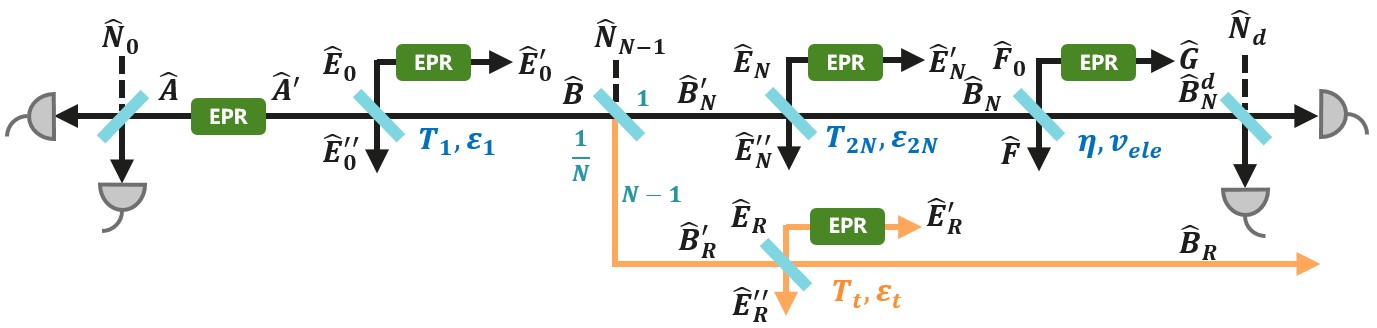}
        \caption{\label{SuppSimpEB2User}
        \textbf{The 2-user EB scheme for security analysis.}}
    \end{figure}
    This part is a detailed security analysis with the 2-user scheme, corresponding to the simulation results in main text.
    The 2-user EB scheme is shown in Fig. \ref{SuppSimpEB2User}, Eve introduces no correlations between the quantum channels. Here, the quantum channels of the network including three quantum channels: A channel connecting Alice and the beam splitter together with channel parameters $(T_1,\varepsilon_1)$, called $Ch_0$. A channel connecting the beam splitter and Bob $N$ together with parameters $(T_{2N},\varepsilon_{2N})$, called $Ch_N$. The equivalent one-way channel where the output mode is $\hat{B}_R$ and the equivalent channel parameters are $(T_t,\varepsilon_t)$, called $Ch_R$.  Here, the transmittance of the beam splitter is based on the number of Bobs in the network, when there are $N$ Bobs in the network, the transmittance is set as $\frac{N-1}{N}$.
    
    Based on the transformation of the modes, after passing the first channel, mode $A'$ is transformed to mode $\hat{B}$, thus
    \begin{equation}
        \hat{B} = \sqrt{T_1} \hat{A}' +\sqrt{1-T_1}\hat{E}_0.
    \end{equation}
    Then, mode $B$ passes the beam splitter with transmittance $1/N$, 
    \begin{equation}
        \hat{B}'_N= \sqrt{\frac{1}{N}}\hat{B}+\sqrt{1-\frac{1}{N}}\hat{N}_{N-1}.
    \end{equation}
    \begin{equation}
        \hat{B}'_R = \sqrt{1-\frac{1}{N}}\hat{N}_{N-1}-\sqrt{\frac{1}{N}}\hat{B}.
    \end{equation}
    After that, mode $\hat{B}'_N$ and mode $\hat{B}'_R$ are transform to $\hat{B}_N$ and $\hat{B}_R$ respectively,
    \begin{equation}
        \hat{B}_N = \sqrt{T_{2N}}\hat{B}'_N+\sqrt{1-T_{2N}}\hat{E}_N,
    \end{equation}
    \begin{equation}
        \hat{B}_R = \sqrt{T_t}\hat{B}'_R+\sqrt{1-T_t}\hat{E}_R.
    \end{equation}
    Therefore, the modes $\hat{B}_N$ and $\hat{B}_R$ can be written as 
    \begin{equation}
        \begin{split}
            \hat{B}_N &= \sqrt{T_{2N}}(\sqrt{\frac{1}{N}}\hat{B}+\sqrt{1-\frac{1}{N}}\hat{N}_{N-1})+\sqrt{1-T_{2N}}\hat{E}_N\\
            &=\sqrt{T_{2N}}(\sqrt{\frac{1}{N}}(\sqrt{T_1} \hat{A}' +\sqrt{1-T_1}\hat{E}_0)+\sqrt{1-\frac{1}{N}}\hat{N}_{N-1})+\sqrt{1-T_{2N}}\hat{E}_N\\
            &=\sqrt{\frac{T_1T_{2N}}{N}}\hat{A}'+\sqrt{\frac{(1-T_1)T_{2N}}{N}}\hat{E}_0+\sqrt{T_{2N}(1-\frac{1}{N})}\hat{N}_{N-1}+\sqrt{1-T_{2N}}\hat{E}_N.
        \end{split}
    \end{equation}
    \begin{equation}
        \begin{split}
            \hat{B}_R&=\sqrt{T_t}\hat{B}_R^\prime+\sqrt{1-T_t}\hat{E}_R\\
            &=\sqrt{T_t}\left(\sqrt{\frac{1}{N}}\hat{N}_{N-1}-\sqrt{1-\frac{1}{N}}\left(\sqrt{T_1}\hat{A}^\prime+\sqrt{1-T_1}\hat{E}_0\right)\right)+\sqrt{1-T_t}\hat{E}_R\\
            & = \sqrt{\frac{T_t}{N}}\hat{N}_{N-1}-\sqrt{T_1T_t\left(1-\frac{1}{N}\right)}\hat{A}'-\sqrt{\left(1-T_1\right)T_t\left(1-\frac{1}{N}\right)}\hat{E}_0+\sqrt{1-T_t}\hat{E}_R.
        \end{split}
    \end{equation}
    As usual, the detector is trusted and modeled as one always do in the security analysis of continuous-variable quantum key distribution, the mode $\hat{B}_N$ is transformed to mode $\hat{B}_N^d$
    \begin{equation}
        \begin{split}
            \hat{B}_N^d&=\sqrt\eta \hat{B}_N+\sqrt{1-\eta}\hat{F}_0\\
            &=\sqrt\eta (\sqrt{\frac{T_1T_{2N}}{N}}\hat{A}'+\sqrt{\frac{(1-T_1)T_{2N}}{N}}\hat{E}_0+\sqrt{T_{2N}(1-\frac{1}{N})}\hat{N}_{N-1}+\sqrt{1-T_{2N}}\hat{E}_N)+\sqrt{1-\eta}\hat{F}_0\\
            &=\sqrt{\frac{\eta T_1T_{2N}}{N}}\hat{A}^\prime+\sqrt{\frac{\eta\left(1-T_1\right)T_{2N}}{N}}\hat{E}_0+\sqrt{\eta\left(1-T_{2N}\right)}\hat{E}_N+\sqrt{1-\eta}\hat{F}_0+\sqrt{\eta T_{2N}\left(1-\frac{1}{N}\right)}\hat{N}_{N-1}.
        \end{split}
    \end{equation}
    The other output modes are $\hat{F}$ and $\hat{G}$, mode $\hat{G}$ has no correlation besides mode $\hat{F}$, 
    \begin{equation}
        \begin{split}
            \hat{F}&=\sqrt\eta \hat{F}_0-\sqrt{1-\eta}\hat{B}_N\\
            &=\sqrt\eta  \hat{F}_0-\sqrt{\frac{\left(1-\eta\right)T_1T_{2N}}{N}} \hat{A}^\prime-\sqrt{\frac{\left(1-\eta\right)\left(1-T_1\right)T_{2N}}{N}} \hat{E}_0-\sqrt{(1-\eta)(1-\frac{1}{N})} \hat{N}_{N-1}+\sqrt{1-T_{2N}} \hat{E}_N.
        \end{split}
    \end{equation}
    
    According to the mode transformation relationship above, the variance of each modes and the covariance between different modes can be obtained.
    
    For mode $\hat{A}$: The variance of mode $\hat{A}$, $V_{{A}}$, is as below, here $V_M$ represents the modulation variance.
    \begin{equation}
        V_{{A}} = V_M+1.
    \end{equation}
    The covariance between mode $\hat{A}$ and mode $\hat{B}_N$ can be written as $C_{{A}{B}_N}$
    \begin{equation}
        C_{{A}{B}_N}=\sqrt{\frac{T_1T_{2N}}{N}}C_{AA^\prime}=\sqrt{\frac{T_1T_{2N}}{N}}\sqrt{V_A^2-1}.
    \end{equation}
    The covariance between mode $\hat{A}$ and mode $\hat{B}_R$ can be written as $C_{{A}{B}_R}$
    \begin{equation}
        C_{{A}{B}_R}=-\sqrt{{T_1T}_t\left(1-\frac{1}{N}\right)}C_{{A}{A}^\prime}=-\sqrt{{T_1T}_t\left(1-\frac{1}{N}\right)}\sqrt{V_{{A}}^2-1}.
    \end{equation}
    
    For mode $\hat{B}_N$: The variance of mode $\hat{B}_N$, $V_{{B}_N}$, is as below.
    \begin{equation}
        \begin{split}
            V_{{B}_N}&=\frac{T_1T_{2N}}{N}V_{{A}^\prime}+\frac{\left(1-T_1\right)T_{2N}}{N}V_{{E}_0}+\left(1-T_{2N}\right)V_{{E}_N}+T_{2N}\left(1-\frac{1}{N}\right)\\
            &=\frac{T_1T_{2N}}{N}\left(V_M+1\right)+\frac{T_{2N}}{N}\left(1-T_1+T_1\varepsilon_1\right)+1-T_{2N}+T_{2N}\varepsilon_{2N}+T_{2N}\left(1-\frac{1}{N}\right)\\
            &=\frac{T_1T_{2N}}{N}V_M+\frac{T_1T_{2N}}{N}+\frac{T_{2N}}{N}-\frac{T_{2N}T_1}{N}+\frac{T_{2N}T_1\varepsilon_1}{N}+1-T_{2N}+T_{2N}\varepsilon_{2N}+T_{2N}-\frac{T_{2N}}{N}\\
            &=\frac{T_1T_{2N}}{N}V_M+\frac{T_{2N}T_1\varepsilon_1}{N}+1+T_{2N}\varepsilon_{2N}\\
            &=\frac{T_1T_{2N}}{N}\left(V_M{+\varepsilon}_1+\frac{N}{T_1}\varepsilon_{2N}\right)+1\\
            &=\frac{T_1T_{2N}}{N}\left(V_M+\varepsilon_{tot}\right)+1.
        \end{split}
    \end{equation}
    Here,  
    \begin{equation}
        \label{epstot}
        \varepsilon_{tot}=\varepsilon_1+\frac{N}{T_1}\varepsilon_{2N}.
    \end{equation}
    $\varepsilon_{tot}$ represents the excess noise at the Alice side, at the start of $Ch_0$. $\varepsilon_1$ and $\varepsilon_{2N}$ are the excess noise of $Ch_0$ and $Ch_N$. The rationality of setting the excess noise as above is explained in next part. 
    The covariance between mode $\hat{B}_N$ and $\hat{B}_R$, $C_{{B}_R{B}_N}$ is 
    \begin{equation}
        \begin{split}
            C_{{B}_R{B}_N}&=-\sqrt{{T_1T}_t\left(1-\frac{1}{N}\right)}\sqrt{\frac{T_1T_{2N}}{N}}V_{{B}_0}+\sqrt{\frac{T_t}{N}}\sqrt{T_{2N}\left(1-\frac{1}{N}\right)}-\sqrt{\left(1-T_1\right)T_t\left(1-\frac{1}{N}\right)}\sqrt{\frac{\left(1-T_1\right)T_{2N}}{N}}V_{\hat{E}_0}\\
            &=-\frac{T_1}{N}\sqrt{T_{2N}T_t\left(N-1\right)}V_{{B}_0}+\frac{1}{N}\sqrt{{T_{2N}T}_t\left(N-1\right)}-\frac{\left(1-T_1\right)}{N}\sqrt{{T_{2N}T}_t\left(N-1\right)}V_{{E}_0}\\
            &=\frac{-T_1}{N}\sqrt{{T_{2N}T}_t\left(N-1\right)}\left(V_M+\varepsilon_1\right).
        \end{split}
    \end{equation}
    
    For mode $\hat{B}_R$: The variance of mode $\hat{B}_R$, $V_{{B}_R}$, is as below.
    \begin{equation}
        \begin{split}
            V_{{B}_R}&=\frac{T_t}{N}+{T_1T}_t\left(1-\frac{1}{N}\right)\left(V_M+1\right)+T_t\left(1-\frac{1}{N}\right)\left(1-T_1+T_1\varepsilon_1\right)+1-T_t+T_t\varepsilon_t\\
            &={T_1T}_tV_M-\frac{1}{N}{T_1T}_tV_M+T_1T_t\varepsilon_1-\frac{T_1T_t\varepsilon_1}{N}+1+T_t\varepsilon_t\\
            &=\left(1-\frac{1}{N}\right){T_1T}_tV_M+\left(1-\frac{1}{N}\right)T_1T_t\varepsilon_1+1+T_t\varepsilon_t\\
            &=\left(1-\frac{1}{N}\right){T_1T}_t\left(V_M+\varepsilon_{tot}^\prime\right)+1.
        \end{split}
    \end{equation}
    Here, 
    \begin{equation}
        \varepsilon_{tot}^\prime=\varepsilon_1+\frac{\varepsilon_t}{\left(1-\frac{1}{N}\right)T_1}.
    \end{equation}
    When $N\rightarrow \infty$,
    \begin{equation}
        \varepsilon_{tot}^\prime=\varepsilon_1+\frac{\varepsilon_t}{T_1}.
    \end{equation}
    Compared with eq. \ref{epstot}, the coefficient $N$ is missing here, which means that the equivalent one-way parameter $\varepsilon_t$ has little impact on performance of the network. For secret key rate of Bob $N$, the characteristics of his own  channel has the greatest impact.
    
    With the variance and covariance of the modes as mentioned before, the matrix $\gamma_{{A}{B}_R{B}_N}$ which represents the characteristics of the network can be constructed.
    \begin{equation}
        \gamma_{{A}{B}_R{B}_N}=
    \begin{pmatrix}
      \gamma_{{A}}&  \gamma_{{A}{B}_R} & \gamma_{{A}{B}_N} \\
      \gamma_{{A}{B}_R}& \gamma_{{B}_R} & \gamma_{{B}_R{B}_N}\\
      \gamma_{{A}{B}_N}& \gamma_{{B}_R{B}_N} & \gamma_{{B}_N}
    \end{pmatrix}.
    \end{equation}
    In the matrix above, 
    \begin{equation}
        \gamma_{{A}}=
        \begin{pmatrix}
            V_{{A}} & 0 \\
            0 & V_{{A}}
        \end{pmatrix}
        =(V_M+1)\cdot I_2.
    \end{equation}
    \begin{equation}
        \gamma_{{A}{B}_R}=
        \begin{pmatrix}
            C_{{A}{B}_R} & 0\\
            0 & -C_{{A}{B}_R}\\
        \end{pmatrix}
        =-\sqrt{{T_1T}_t\left(1-\frac{1}{N}\right)}\sqrt{V_{{A}}^2-1}\cdot \sigma_z.
    \end{equation}
    \begin{equation}
        \gamma_{{A}{B}_N}=
        \begin{pmatrix}
            C_{{A}{B}_N} & 0\\
            0 & -C_{{A}{B}_N}\\
        \end{pmatrix}
        =\sqrt{\frac{T_1T_{2N}}{N}}\sqrt{V_A^2-1} \cdot \sigma_z.
    \end{equation}
    \begin{equation}
        \gamma_{{B}_R}=
        \begin{pmatrix}
            V_{{B}_R} & 0 \\
            0 & V_{{B}_R}
        \end{pmatrix}
        =\left(\left(1-\frac{1}{N}\right){T_1T}_t\left(V_M+\varepsilon_{tot}^\prime\right)+1\right)\cdot I_2.
    \end{equation}
    \begin{equation}
        \gamma_{{B}_R{B}_N}=
        \begin{pmatrix}
            C_{{B}_R{B}_N} & 0\\
            0 & C_{{B}_R{B}_N}\\
        \end{pmatrix}
        =\frac{-T_1}{N}\sqrt{{T_{2N}T}_t\left(N-1\right)}\left(V_M+\varepsilon_1\right)\cdot I_2.
    \end{equation}
    \begin{equation}
        \gamma_{{B}_N}=
        \begin{pmatrix}
            V_{{B}_N} & 0 \\
            0 & V_{{B}_N}
        \end{pmatrix}
        =\left(\frac{T_1T_{2N}}{N}\left(V_M+\varepsilon_{tot}\right)+1 \right)\cdot I_2.
    \end{equation}
    Here, 
    $I_2=\begin{pmatrix}
        1&0\\0&1
    \end{pmatrix}$,
    $\sigma_z=\begin{pmatrix}
        1&0\\0&-1
    \end{pmatrix}$. 
    With the matrix $\gamma_{{A}{B}_R{B}_N}$, the matrix with $\hat{B}_N$ trusted modeled is 
    \begin{equation}
        \gamma_{{A}{B}_R{B}_N^d{F}{G}}=(Y^{BS})^T[\gamma_{{A}{B}_R{B}_N}\oplus \gamma_{{F}_0{G}}]Y^{BS}.
    \end{equation}
    Here, 
    \begin{equation}
        \gamma_{{F}_0{G}}=
        \begin{pmatrix}
            v\cdot I_2 & \sqrt{(v^2-1)}\cdot\sigma_z\\
            \sqrt{(v^2-1)}\cdot\sigma_z & v\cdot I_2
        \end{pmatrix}.
    \end{equation}
    For heterodyne detection, $v=1+\frac{2 v_{el}}{1-\eta_d}$, $v_{el}$ is the electronic noise of the balanced homodyne detector, $\eta_d$ is the detection efficiency of the detector.
    By adjusting the modes, $\gamma_{{A}{B}_R{F}{G}{B}_N^d}$ can be calculated. Then we can get,
    \begin{equation}
        \gamma_{{A}{B}_R{F}{G}}^{m_{{B}_N^d}}=\gamma_{{A}{B}_R{F}{G}}-\sigma_{{A}{B}_R{F}{G}{B}_N^d}^T H \sigma_{{A}{B}_R{F}{G}{B}_N^d}.
    \end{equation}
    For heterodyne detection, $H=(\gamma_{{B}_N^d}+I_2)^{-1}$. $\gamma_{{A}{B}_R{F}{G}}$, $\sigma_{{A}{B}_R{F}{G}{B}_N^d}$ and $\gamma_{{B}_N^d}$ can be derived from the decomposition of $\gamma_{{A}{B}_R{F}{G}{B}_N^d}$ as below
    \begin{equation}
        \gamma_{{A}{B}_R{F}{G}{B}_N^d}=
        \begin{pmatrix}
            \gamma_{{A}{B}_R{F}{G}} & \sigma_{{A}{B}_R{F}{G}{B}_N^d}^T \\
            \sigma_{{A}{B}_R{F}{G}{B}_N^d} & \gamma_{{B}_N^d}
        \end{pmatrix}.
    \end{equation}
    The symplectic eigenvalues of the matrix $\gamma_{{A}{B}_R{F}{G}{B}_N^d}$ and $\gamma_{{A}{B}_R{F}{G}}^{m_{{B}_N^d}}$ representing the characteristics of state $\rho_{{A}{B}_R{F}{G}{B}_N^d}$ and $\rho_{{A}{B}_R{F}{G}}^{m_{{B}_N^d}}$ can be calculated. The symplectic values of the two matrix larger than 1 can be written as $\lambda_{i}$ and $\lambda'_{j}$.
    The Von Neumann entropy of the quantum state $\rho$ can be written as $S(\rho)$, and
    \begin{equation}
        S(\rho_{{A}{B}_R{F}{G}{B}_N^d})=\sum_{i}G(\frac{\lambda_i-1}{2}),
    \end{equation}
    \begin{equation}
        S(\rho_{{A}{B}_R{F}{G}}^{m_{{B}_N^d}})=\sum_{j}G(\frac{\lambda'_j-1}{2}).
    \end{equation}
    Here, $G(x)=(x+1)log_2(x+1)-xlog_2x$. 
    Assuming that Eve can purifies the system, with the extremality of Gaussian states \cite{wolf2006extremality}, the Holevo bound $\chi_{B_NE}$ \cite{holevo1973bounds} which represents the upper bound of the potential information being eavesdropped can be calculated by 
    \begin{equation}
        \chi_{{B}_N{E}} = S(\rho_{{A}{B}_R{F}{G}{B}_N^d}) - S(\rho_{{A}{B}_R{F}{G}}^{m_{{B}_N^d}}).
    \end{equation}
    Thus, the secret key rate of the network with a simplified scheme is
    \begin{equation}
        K_N = \beta I\left({A}:{B}_N\right)-\max{\left\{I^{max}_{{B}_N{B}},\chi_{{B}_N{E}}\right\}}.
    \end{equation}
    Here, $\beta$ is the reconciliation efficiency, $\beta I(A:B_N)$ is the information that can be used by Alice and Bob $N$, 
    \begin{equation}
        \beta I\left({A}:{B}_N\right)=\beta log_2((V_{{A}}+1)/(V_{{A}|{B}_{Nx}}+1)).
    \end{equation}
    $I^{max}_{{B}_N{B}}$ is the maximum value of the mutual information between different Bobs. Because the Bobs are the legitimate parties, thus their correlation can be calculated with classical mutual information.
    \begin{equation}
        \begin{split}
            I^{max}_{{B}_N{B}}=\max\{I\left({B}_N:{B}_1\right),\ I\left({B}_N:{B}_2\right),\ldots,&I\left({B}_N:{B}_{N-1}\right)\},
        \end{split}
    \end{equation}
    \begin{equation}
        I ({B}_N:{B}_i)=log_2(V_{{B}_{N_x}^d}/V_{{B}_{N_x}^d|{B}_{i_x}}).
    \end{equation}
    
    In this way, the secret key rate of the 2-user EB scheme of the PTMP quantum secure network can be calculated. This can be used for simplifying the security analysis in simulation of the network when assuming there are no correlation introduced by Eve between the quantum channels. Especially when the quantum channels connecting the optical power splitter and the users have the same channel parameters, the simplification can be very simple, and there are no loss of the secret key rate due to the simplification.
    
    \section{Supplementary Note 4: The setting of the simulation parameters}
    For simulation of the network, in the EB scheme, the channel parameters of the channels should be settled. These channels can be divided into two categories, the first is the channel connects Alice with the optical power splitter, the excess noise of this channel, $\varepsilon_1$ in Fig. \ref{SuppSimpEB2User}, represents the noise which has the same impact on all of the Bobs. The channel which connects the optical power splitter with Bobs is the second one, represents the noise which has different impact on Bobs, written as $\varepsilon_N$ and $\varepsilon_R$ in Fig. \ref{SuppSimpEB2User}. Here, $\varepsilon_R$ represents the noise from the other links, and $\varepsilon_N$ represents the noise of the channel of Bob $N$, whose secret key rate is being calculated.
    
    In the network where no eavesdropper is actually performing attack, the noise mainly consists of the detection noise, the Raman noise, the phase recovery noise, the analog to digital converter (ADC) noise, and the source noise \cite{Noise}. Here, the Raman noise mainly originates from the coexistence of the quantum signals and classical signals, for simulation of the performance of the protocol, it can be ignored. The ADC noise, phase recovery noise and the detection noise all acts on the detection data, which has different impact on different Bobs. Thus, these noise should be added to the channel linking the optical power splitter and the receivers of Bobs. In the 2-user scheme, these noise are $\varepsilon_{2N}$ and $\varepsilon_{t}$.  Especially, when the detector is trusted modeled, the detection noise is removed from $\varepsilon_{2N}$ and $\varepsilon_{t}$. Here, the main source of the noise $\varepsilon_{2N}$ and $\varepsilon_{t}$. is the phase recovery noise.
    The noise from the source mainly originates from the modulation error, which has the same impact on all of the Bobs, thus, this noise should be added to the channel linking Alice and the optical power splitter, which is $\varepsilon_{1}$.
    
    It's hard to define a total excess noise of the network, because in the security analysis of the network scheme, the channel is divided by a power splitter.
    To set a reasonable excess noise value for the channels, as well as to make the excess noise in the network correspond to that of the one-way system , a model of excess noise is used. Based on the impact of the excess noise on quantum states mentioned above, $\varepsilon_1$ mainly comes from the noise of the source, and $\varepsilon_N$ mainly comes from the phase recovery noise of the receivers. In the respect of the noise model, the modulation noise from the source and the phase recovery noise at the receiver can be written as 
    \begin{equation}
        \begin{split}
            \xi_{mod} &= TV_{mod}\delta_{mod}, \\
            \xi_{PR} &= TV_{mod}\delta_{PR}.
        \end{split}
    \end{equation}
    here, $\xi$ represents the equivalent excess noise at the receiver, $T$ is the transmittance of the whole channel, $V_{mod}$ is the modulation variance of the OLT, $\delta_{mod}$ and $\delta_{PR}$ are the fixed error parameters of modulation and phase recovery. This is the noise model based on the source of the noise in physical layer, thus it is still applicable in the network.
    Assume that the transmittance of the feeder fiber is $T_1$, the transmittance of all the drop fibers are the same, which is $T_2$, and the power splitter introduce a loss of $1/N$, where $N$ is the number of the users. 
    In this way, we can define the total excess noise of one point to point link of the network as below, which can be compared with the one-way protocol, 
    Thus, we define the equivalent total excess noise at the input of the feeder fiber as 
    \begin{equation}
        \varepsilon_{tot} = V_{mod}(\delta_{mod}+\delta_{PR}) = \varepsilon_{mod} + \varepsilon_{PR}^{Tx}. 
    \end{equation}
    By comparing with the one-way experiments, the excess noise parameters can be reasonably settled. 
    What's more, with $ \varepsilon_{tot}$ and $L_{tot}$ settled, if the drop fibers are same, the location of the power splitter has no impact on the covariance matrix $\gamma_{AB_1B_2\ldots B_N}$.
\end{widetext}

\end{document}